\documentclass[12pt]{iopart}

\usepackage{graphicx}
\usepackage{iopams} 
\usepackage{prettyref}
\usepackage{feynmp-auto}
\usepackage{comment}
\usepackage{hyperref}
\usepackage{setstack}
\usepackage{xcolor}

\newrefformat{cap}{\hyperref[#1]{Figure~\ref{#1}}}
\newrefformat{fig}{\hyperref[#1]{Figure~\ref{#1}}}
\newrefformat{tab}{\hyperref[#1]{Table ~\ref{#1}}}
\newrefformat{sec}{\hyperref[#1]{Section~\ref{#1}}}
\newrefformat{sub}{\hyperref[#1]{Section~\ref{#1}}}
\newrefformat{cha}{\hyperref[#1]{Chapter~\ref{#1}}}
\newrefformat{app}{\hyperref[#1]{Appendix~\ref{#1}}}

\begin{document}


\newcommand{\cH}{\mathscr{H}}
\newcommand{\C}{\mathscr{C}}
\newcommand{\dd}{{\rm d}}
\newcommand{\sR}{\text{\tiny R}}
\newcommand{\ee}{{\rm{e}}}
\newcommand{\thy}{\text{th}}
\newcommand{\ji}{\text{\small i}}
\newcommand{\p}{\partial}
\newcommand{\bx}{\text{\bf x}}
\newcommand{\bR}{\text{\bf R}}
\newcommand{\by}{\text{\bf y}}
\newcommand{\bz}{\text{\bf z}}
\newcommand{\be}{\text{\bf e}}
\newcommand{\eps}{\varepsilon}
\newcommand{\bk}{\text{\bf k}}
\newcommand{\bo}{\text{\bf 0}}
\newcommand{\bq}{\text{\bf q}}
\newcommand{\bp}{\text{\bf p}}
\newcommand{\bc}{\bf c}
\newcommand{\bP}{\text{\bf P}}
\newcommand{\bu}{\text{\bf u}}
\newcommand{\bv}{\text{\bf v}}
\newcommand{\ii}{{\rm int}}

\newcommand{\KS}{\text{\tiny KS}}
\newcommand{\kb}{k_\text{\tiny B}}


\global\long\def\T{\mathrm{T}}
\global\long\def\del{\delta}
\global\long\def\Gs{\Gamma}
\global\long\def\SI{\int}
\global\long\def\Gstz{\tilde{\Gamma}_{2}^{0}}%
\global\long\def\km{\kappa_{1}}%
\global\long\def\kv{\kappa_{2}}%
\global\long\def\ks{\kappa_{3}}%
\global\long\def\kk{\kappa_{4}}%
\global\long\def\J{{\it M}}%

\title{Diagrammatics for the Inverse Problem in Spin Systems and Simple Liquids}

\author{Tobias K\"uhn}
\address{Laboratoire Matière et Systèmes Complexes, Université Paris Cité \& CNRS (UMR 7057), 10 rue Alice Domon et Léonie Duquet, 75013 Paris, France}
\address{Institut de la Vision, Sorbonne Université \& INSERM \& CNRS (UMR S968), 17 rue Moreau, 75012, Paris, France}

\ead{tobias.kuhn@inserm.fr}

\author{Fr\'ed\'eric van Wijland}

\address{Laboratoire Matière et Systèmes Complexes, Université Paris Cité \& CNRS (UMR 7057), 10 rue Alice Domon et Léonie Duquet, 75013 Paris, France}

\begin{abstract}
Modeling complex systems, like neural networks, simple liquids or flocks of birds, often works in reverse to textbook approaches: given data for which averages and correlations are known, we try to find the parameters of a given model consistent with it. In general, no exact calculation directly from the model is available and we are left with expensive numerical approaches. A particular situation is that of a perturbed Gaussian model with polynomial corrections for continuous degrees of freedom. Indeed perturbation expansions for this case have been implemented in the last 60 years. However, there are models for which the exactly solvable part is non-Gaussian, such as independent Ising spins in a field, or an ideal gas of particles. We implement a diagrammatic perturbative scheme in weak correlations around a non-Gaussian yet solvable probability weight. This applies in particular to spin models (Ising, Potts, Heisenberg) with weak couplings, or to a simple liquid with a weak interaction potential. Our method casts systems with discrete degrees of freedom and those with continuous ones within the same theoretical framework. When the core theory is Gaussian it reduces to the well-known Feynman diagrammatics.

\end{abstract}

%
%
%
%
%

\section{Motivations}

Consider a many-body problem, such as one involving interacting spins or particles, for which the available data, whether experimental or numerical, allows us to determine the corresponding averages and two-point correlation functions. Suppose the distribution of the spin values or of the particle positions is written in a Boltzmann form with one- and two-body interactions only. One  question we wish to adress is how to choose the corresponding couplings so that the distribution generates the correct two-point correlations. Such inverse problems are of course ubiquitous in the study of complex systems from simple fluids to neurons or proteins (see for instance \cite{henderson1974uniqueness,mackay2003information,schneidman2006weak,Sessak09,Jacquin2016resummed,cocco2018inverse,campos2021machine}). In general, making analytical progress for inverse problems requires to resort to approximation schemes. Even when only the averages of the degrees of freedom are constrained this is a formidable task equivalent to solving a statistical mechanical model with a one-body external field. Outside the realm of Gaussian theories which are exactly solvable, few generic methods exist~\cite{Vasiliev74,Vasiliev75,plefka1982convergence,georges1991expand,Jacquin2016resummed,Kuehn18_375004}. If one further constrains correlations to be fixed, the problem has been adressed at the field-theoretic level (that rests on an expansion around a Gaussian theory) and is known under various names: Luttinger and Ward, or 2PI, or effective potential methods~\cite{berges2004introduction,calzetta2008nonequilibrium}. Within the realm of simple liquids, the Poisson nature of statistics of the ideal gas has also been exploited~\cite{morita1960new,morita1961new,vasiliev2019functional} to derive a one-to-one connection between the pair potential and the pair distribution function. For Ising spins (binary variables), weak correlation expansions have been carried out as well~\cite{Sessak09}. However, it is our feeling that, however remarkable, these inroads into specific systems, such as simple liquids or Ising spins, could benefit a more general approach, applicable, {\it e.g.} to Potts spins or even, more speculatively, to trajectory-dependent observables~\cite{andreanov2006dynamical,kim2014equilibrium,bravi2016extended}.\\

We begin by introducing the framework designed to work at given two-point correlations in \prettyref{sec:2PIfunctional}, by introducing a Legendre transform with respect to correlations, just as is used in the Luttinger-Ward approach. In that sense we build on the work of \cite{Kuehn18_375004} where the authors focused on the Legendre transform with respect to averages. This leads us to defining a free-energy like functional of the two-point function, which attains its minimum at the prescribed physical two-point correlation function. We then outline our perspective on the diagrammatic expansion of this technique around Gaussian theories in \prettyref{sec:Gaussian_theory_main}, reproducing the classical results for this case. The main novelty of this work is to show that a related diagrammatic expansion of this functional in the correlations is possible, provided the one-body problem is solvable, though not necessarily characterized by a Gaussian distribution. Equipped with this tool, we then implement the procedure on the Ising model, where the Sessak-and-Monasson results~\cite{Sessak09} are readily recovered, and further extended to $q$-state variables ($q\geq 2$). Besides that, we show that our approach is consistent with the Hiroike-and-Morita expansion~\cite{morita1960new} for simple liquids, and field-theoretic approaches based on the Luttinger-Ward functional. In our conclusion we suggest interesting applications.

\section{The second Legendre transform}\label{sec:2PIfunctional}
Let $H\left[\psi\right]$ be a dimensionless Hamiltonian for a field $\psi_{x}$, where $\psi_x$ and/or $x$ can be living in a discrete or continuous space (for a discrete $\psi$, $\int\dd\psi\ldots$ refers to a summation over the allowed values of $\psi$). For this Hamiltonian, we define the cumulant-generating functional
\begin{eqnarray}
	\label{eq:Def_cumulant_generating_fct}
	W\left[j,K\right]=\ln \int \dd\psi\,\ee^{-H\left[\psi\right]+ j^\T \psi+\frac{1}{2}\psi^{\T}K\psi},
\end{eqnarray}
which depends not only on a source term $j$ for the field $\psi$, but also on a symmetric source $K$ coupled to a two-point observable. The sources $j$ and $K$ live in the same $x$-space as the field $\psi$, and we shall resort to the notational shortcut
\begin{eqnarray}
	j^\T\psi=\int_{x} j_{x}\psi_{x},\quad\psi^\T K\psi=\int_{x,y}\,\psi_{x}K_{x,y}\psi_{y},
\end{eqnarray}
where the integration sign is replaced with a discrete summation depending on the discrete/continuous nature of $x$. Since we are interested in investigating an inverse problem, where the first two moments of $\psi$ are prescribed, we find it convenient to Legendre transform $W$ with respect to both sources. Consider the functional $\Gamma$ depending on a one-body field $\phi$ and on a symmetric two-body field $\Delta$ defined by 
\begin{eqnarray}
	\Gamma\left[\phi,\Delta\right] = \sup_{j,K} \left[j^{\T}\phi+\frac{1}{2}\mathrm{tr}\left[K\left(\Delta+\phi \phi^{\T}\right)\right] - W\left[j,K\right]\right].\label{eq:Def_second_Legendre}
\end{eqnarray}
In particle and condensed-matter physics~\cite{berges2004introduction,calzetta2008nonequilibrium}, this is known as the Luttinger-Ward or Baym–Kadanoff functional or 2PI functional, in liquid-state theory as the Morita-Hiroike functional and in statistics as maximum log-likelihood (at given second-order statistics). In the remainder, we will refer to it as the effective potential and denote it by $\Gs$. 
At the supremum in expression \prettyref{eq:Def_second_Legendre} we have
\begin{equation}
	\phi_{x} = \frac{\delta W}{\delta j_x},\,\, \Delta_{xy} + \phi_{x}\phi_{y}=\frac{\delta W}{\delta K_{xy}}\label{eq:def_Delta_deriv_j_K}.
\end{equation}
The functional derivatives of $\Gamma$ are found to be given by
\begin{equation}
	\label{eq:Eq_of_state}
	\frac{\delta\Gamma}{\delta\phi_x}=j_x+\int_{y} K_{xy} \phi_y,\,\frac{\delta \Gamma}{\delta \Delta_{xy}}=K_{xy}
\end{equation}
and in the physically relevant limit of vanishing sources $j$ and $K$, these \prettyref{eq:Eq_of_state} make up the equations of state in the sense that they determine the values of the one and two-point correlations, $\phi_x=\langle\psi_x\rangle$ and $\Delta=\langle\psi_x\psi_y\rangle-\langle\psi_x\rangle\langle\psi_y\rangle$: The functional $\Gs$ is minimized when $\phi$ and $\Delta$ are, respectively, the one-point and connected two-point correlation functions. These relations tell us that if we were able to determine $\Gs$ exactly, then by performing the corresponding functional derivatives in \prettyref{eq:Eq_of_state}, we would solve the inverse problem. This is of course a very difficult problem due to the presence of interactions. The idea is thus to separate the one-body problem, which we assume to be solvable (in the sense that the corresponding effective potential can be found exactly), and to expand the functional in powers of the correlations induced by two-body interactions.\\
To prepare for the implementation of this expansion, we derive an expression for $\Gs$ without explicit reference to the Legendre transform. We use the definition of $\Gs$, Eq.~\prettyref{eq:Def_second_Legendre}, with the conditions Eqs.~\prettyref{eq:def_Delta_deriv_j_K} on $j$ and $K$, to arrive at
\begin{eqnarray}
	\fl &\int d\psi\,\exp\left(-H\left[\psi\right]+j^{\T}\psi+\frac{1}{2}\psi^{\T}K\psi\right) = \exp\left(W\left[j,K\right]\right)\\ 
	\fl = &\exp\left(\phi^{\T} j+\frac{1}{2}\mathrm{tr}\left(K\left(\Delta+\phi\phi^{\T}\right)\right)-\Gs\left[\phi,\Delta\right]\right)\\
 	\fl = &\exp\left(\phi^{\T}\left(\frac{\del\Gs}{\del\phi}-\frac{\del\Gs}{\del\Delta}\phi\right)+\frac{1}{2}\mathrm{tr}\frac{\del\Gs}{\del\Delta}\left(\Delta+\phi\phi^{\T}\right)-\Gs\left[\phi,\Delta\right]\right),
\end{eqnarray}
where we have made use of Eq.~\prettyref{eq:Eq_of_state} to express the source terms $j$ and $K$ in terms of derivatives of $\Gs$ and we have implied the summation over indices. We finally isolate $\Gs$ and replace the field $\psi$ with its expression as a derivative with respect to $j$. This leads to
\begin{eqnarray}\label{eq:eqGammaisol}
	\fl & \exp\left(-\Gs\left[\phi,\Delta\right]\right)\\
	\fl = & \int d\psi\,
	\exp\left(-H\left[\psi\right]+\left(\psi-\phi\right)^{\T}\left(\frac{\del\Gs}{\del\phi}-\frac{\del\Gs}{\del\Delta}\phi\right)+\frac{1}{2}\mathrm{tr}\frac{\del\Gs}{\del\Delta}\left[\left(\psi\psi^{\T}-\phi\phi^{\T}\right)-\Delta\right]\right) \nonumber\\
	\fl = & \exp\left(\left(\frac{\del}{\del j} - \phi\right)^{\T}\frac{\del\Gs}{\del\phi} + \frac{1}{2}\mathrm{tr}\frac{\del\Gs}{\del\Delta}\left[\left(\frac{\del}{\del j} - \phi\right) \left(\frac{\del}{\del j} - \phi\right)^{\T} -\Delta\right]\right) \left.\int d\psi\,e^{-H\left[\psi\right]+j^{\T} \psi}\right|_{j=0}.\label{eq:Gamma2_integral_expression}
\end{eqnarray}
At this stage, no approximation has been made, and Eq.~\prettyref{eq:eqGammaisol} is fully exact. The next section sets the stage for the perturbation expansion, and introduces a diagrammatic procedure to perform a weak correlation expansion.

\section{Effective potential in powers of the interaction}

\subsection{A useful identity on the contribution of interactions}
Consider the situation in which $H$ is the sum of a solvable part $H_{0}$ and of a perturbation $H_{\ii}$. 
We take $\Gs^{0}$ to be the first Legendre transform of the unperturbed part. If the unperturbed theory remains solvable upon introducing a two-point source (as is the case for a quadratic $H_0$), we can also choose $\Gs^{0}$ to be its second Legendre transform. We further comment this in appendix \ref{General_non_Gauss}. We then define  $\Gs^{\ii}$ by
\begin{eqnarray}
	\label{eq:Splitting_Gs}
	\Gs\left[\phi, \Delta\right] = \Gs^{0}[\phi] + \Gs^{\ii}\left[\phi, \Delta\right].
\end{eqnarray}
It is important to realize that $\Gs^{0}[\phi]$ does not depend on $\Delta$, as we expect, since it describes a noninteracting theory. We now plug Eq.~\prettyref{eq:Splitting_Gs} into Eq.~\prettyref{eq:Gamma2_integral_expression} and isolate $\Gs^{\ii}$ on the left hand side, which yields
\begin{eqnarray}
	\label{eq:Def_Gamma2V_implicit}
	\fl \exp\left(-\Gamma^{\ii}\left[\phi,\Delta\right]\right)
	=  \left.\exp\left(-W_{0}\left[j\right]\right) \exp\left(A\left[\frac{\del}{\del j}\right]\right) \exp\left(W_{0}\left[j\right]\right)\right|_{j=\frac{\delta\Gs^{0}}{\delta\phi}},
\end{eqnarray}
where
\begin{eqnarray}
	\fl A\left[\frac{\del}{\del j}\right] :=  -H_{\ii}\left[\frac{\del}{\del j}\right]+\frac{\del\Gs^{\ii}}{\del\phi}^{\T}\left(\frac{\del}{\del j}-\phi\right)+\frac{1}{2}\mathrm{tr}\frac{\del\Gs^{\ii}}{\del\Delta}\left[\left(\frac{\del}{\del j}-\phi\right)\left(\frac{\del}{\del j}-\phi\right)^{\T}-\Delta\right].
	\label{eq:Def_A_diff_op}
\end{eqnarray}
To obtain Eq.~\prettyref{eq:Def_Gamma2V_implicit}, we have replaced the term $e^{\Gs^{0}-\frac{\del\Gs^{0}}{\del\phi}\phi}$ with $e^{-W_0}$. The field $j$ is evaluated at $\frac{\delta\Gs^{0}}{\delta \phi}$. This equation is the starting point for the diagrammatics to be developed in the next section.

\subsection{Towards an expression that can be analyzed with diagrams}
In order to proceed with diagrammatics, we will closely follow the footsteps of Kühn and Helias~\cite{Kuehn18_375004}. Indeed, our starting expression Eq.~\prettyref{eq:Def_Gamma2V_implicit} differs from their Eq.~(12) by an additional term due to the two-point source in Eq.~\prettyref{eq:Def_A_diff_op}, but the following steps will be largely analogous. For self-containedness, we carry out---and adapt---the reasoning from scratch. Our approach here is rather formal, and we refer readers familiar to the theory of simple liquids to appendix \ref{sub:First_Mayer_thm} where we illustrate these ideas in the language of Mayer diagrams.  There, we give a proof of the first Mayer theorem (also known as the linked-cluster theorem: $W$ is represented by all connected diagrams) for the example of a simple liquid, using the slightly easier-to-manipulate cumulant-generating function instead of the effective potential.\\

We would like to construct the full theory order by order, using Eq.~\prettyref{eq:Def_Gamma2V_implicit}, so we start by defining the functional $g_{l,L}$ by
\begin{eqnarray}
	e^{g_{l,L}\left[j\right]} := \left(1 + \frac{A\left[\frac{\del}{\del j}\right]}{L} \right)^{l} \exp\left(W_{0}\left[j\right]\right).
\end{eqnarray}
Comparing this expression with Eq.~\prettyref{eq:Def_Gamma2V_implicit}, we obtain $\Gs^{\ii}$ from $g_{l,L}$ via
\begin{eqnarray}
	\fl-\Gs^{\ii}\left[\phi\right] = \left. -W_{0}\left[j\right] + \lim_{L \rightarrow \infty} g_{L,L}\left[j\right]\right|_{j = \frac{\del}{\del \phi}\Gs^{0}\left[\phi\right]} 
	=: \left. -W_{0}\left[j\right] + g\left[j\right]\right|_{j = \frac{\del}{\del \phi}\Gs^{0}\left[\phi\right]}.\label{eq:Gamma_from_g}
\end{eqnarray}
which defines $g$. Using that
\begin{eqnarray}
	e^{g_{l+1,L}\left[j\right]} = \left(1 + \frac{A\left[\frac{\del}{\del j}\right]}{L} \right) e^{g_{l,L}\left[j\right]},
\end{eqnarray}
we derive a recursion relation for $g_{l,L}$:
\begin{eqnarray}
	 \fl & g_{l+1,L}-g_{l,L}\label{eq:Iteration_g} \\
	 \fl = & \frac{1}{L}\,e^{-g_{l,L}\left[j,K\right]}\,\left(-H_{\ii}\left[\frac{\del}{\del j}\right]\right)\,e^{g_{l,L}\left[j,K\right]}\label{eq:iteration_generate_W} \\
 	 \fl  + &\frac{1}{L}\,e^{-g_{l,L}\left[j,K\right]}\,\frac{\del\Gs^{\ii}}{\del\phi}^{\T}\left(\frac{\del}{\del j}-\phi\right)\,e^{g_{l,L}\left[j,K\right]}\label{eq:iteration_1pr_remove} \\
 	 \fl  + &\frac{1}{L}\,e^{-g_{l,L}\left[j,K\right]}\,\frac{1}{2}\mathrm{tr}\frac{\del\Gs^{\ii}}{\del\Delta}\left[\left(\frac{\del}{\del j}-\phi\right)\left(\frac{\del}{\del j}-\phi\right)^{\T}-\Delta\right]\,e^{g_{l,L}\left[j,K\right]}\label{eq:iteration_2pr_remove} \\
 	 \fl  + &\mathcal{O}(L^{-2}),\nonumber
\end{eqnarray}
where $g_{0,L}=W_{0}$ for all $L$. We now give a diagrammatic interpretation to this equation.
\subsection{Constructing a diagrammatic expansion}
To interpret Eq.~\prettyref{eq:Iteration_g} diagrammatically let us first discuss the effect of line \prettyref{eq:iteration_generate_W}. This line tells us that in order to construct the diagrams of order $l$, we have to combine one (bare) interaction with diagrams containing $0$ to $l-1$ interactions. For instance, taking as an example a four-point interaction perturbing a Gaussian theory ($W_{0}$ only contains cumulants of order $1$ and $2$), to first order, we obtain the following diagrams
\begin{fmffile}{First_order_Gauss}
	\fmfset{thin}{0.75pt}
	\fmfset{decor_size}{4mm}
	\begin{eqnarray}\label{eq:Feynmanphi4-1storder}
		\parbox{40mm}{
			\begin{fmfgraph*}(80,30)
				\fmfleft{i1}
				\fmfright{o1}
				\fmf{phantom}{i1,v1,o1}	
				\fmf{plain, tension=.25, right=.5}{v1,i1,v1}
				\fmf{plain, tension=.25, right=.5}{v1,o1,v1}
				\fmfv{d.s=circle, d.filled=empty}{i1,o1}
			\end{fmfgraph*}		
		}	
		+
		\mkern 15mu
		\parbox{30mm}{
			\begin{fmfgraph*}(60,30)
				\fmfleft{i1,i2}
				\fmfright{o1}
				\fmf{phantom}{i1,v1,o2}	
				\fmf{phantom}{i2,v1,o1}
				\fmf{plain}{i1,v1}
				\fmf{plain}{i2,v1}
				\fmf{plain, tension=.25, right=.5}{v1,o1,v1}
				\fmfv{d.s=circle, d.filled=empty}{i1,i2,o1}
			\end{fmfgraph*}		
		}
		\mkern-5mu
		+
		\parbox{20mm}{
			\begin{fmfgraph*}(40,30)
				\fmfleft{i1,i2}
				\fmfright{o1,o2}
				\fmf{plain}{i1,v1,o2}	
				\fmf{plain}{i2,v1,o1}
				\fmfv{d.s=circle, d.filled=empty}{i1,i2,o1,o2}
			\end{fmfgraph*}		
		}.	
	\end{eqnarray}
\end{fmffile}
A circle (node) with $n$ legs represents an $n$-th order cumulant of the unperturbed theory, which are generated by letting the $j$-derivatives acting on $g_{l,L}$ (we recall that constructing the first order $l=1$, $g_{l-1=0,L}$ is identical to $W_{0}$). The crossing of $k$ lines represents a $k$-point interaction.\\

A diagram is translated into a numerical value as follows: Consider that at every leg of the interaction, we have a $j$-derivative acting on $e^{W_{0}\left[j\right]}$. Then we have to count the ways to obtain an expression represented by the first diagram of Eq.~\prettyref{eq:Feynmanphi4-1storder}. The first derivative simply yields $W^{\left(1\right)}\left[j\right]e^{g_{l,L}[j]}$. The second derivative also acts on the prefactor---then the other two together have to generate the other propagator---or it produces another $W^{\left(1\right)}\left[j\right]$-prefactor. In the latter case, we can pick one out of two $W^{\left(1\right)}\left[j\right]$ to let one of the remaining $j$-derivatives act upon. For the last derivative, we have no choice whatsoever. Because all these contributions yield the same value, we have to count the diagram  three times. We will call this the symmetry prefactor of the diagram. It can be determined more conveniently considering the topology of the diagram: Assume that we label the legs of the diagram, say with the indices $i_{1}$ to $i_{4}$. Then, we have three possibilities to choose those legs of the interaction joined by a second-order cumulant. In other words, we obtain the symmetry factor of a diagram by labeling its vertices (and, with them, the cumulants) and count the symmetry operations that leave the unlabeled diagram unchanged, but leads to new labeled diagrams (that is join $\left(i_{1},i_{2}\right)$ and $\left(i_{3},i_{4}\right)$, $\left(i_{1},i_{3}\right)$ and $\left(i_{2},i_{4}\right)$ or $\left(i_{1},i_{4}\right)$ and $\left(i_{2},i_{3}\right)$). So, naming the four-point interaction $U$, we translate
\begin{fmffile}{Example_translation}
	\fmfset{thin}{0.75pt}
	\fmfset{decor_size}{4mm}
	\fl \begin{eqnarray}
		\parbox{40mm}{
			\begin{fmfgraph*}(80,30)
				\fmfleft{i1}
				\fmfright{o1}
				\fmf{phantom}{i1,v1,o1}	
				\fmf{plain, tension=.25, right=.5}{v1,i1,v1}
				\fmf{plain, tension=.25, right=.5}{v1,o1,v1}
				\fmfv{d.s=circle, d.filled=empty}{i1,o1}
			\end{fmfgraph*}		
		}
		\mkern-30mu	
		=
		3\cdot \Delta U \Delta, 
	\end{eqnarray}
\end{fmffile}
where summations over places or momenta are implied.\\

Determining the symmetry factors for the second diagram of the first order in $U$, we need to pick $2$ out of $4$ legs that are connected by a second-order cumulant. We have $\left(\begin{array}{c}4\\2\end{array}\right) = 6$ possibilities to do that, so the symmetry factor is $6$. The derivatives attached to the two legs we have chosen generate the second order cumulant. The remaining ones generate averages (first order cumulants) represented by circles with one outgoing leg each. Generating the third diagram, there is nothing to choose, so the symmetry factor is $1$.\\ 

In diagrams with more than one interaction, in addition, we have to count the possibilities to distribute them. Creating them, the $j$-derivatives might act not only on $W_{0}$, but also on the higher orders of $g_{l,L}$ already generated. By determining the symmetry factor of such a diagram as a whole we can assume that in every iteration step, at most one interaction was added. In this case, one has $\left(\begin{array}{c}L\\ k \end{array}\right)$ possibilities to pick each of these interactions, so in the limit
of $L$ tending to infinity, the prefactor is given by
\begin{eqnarray}
	\fl \lim_{L\rightarrow\infty}\frac{1}{L^{k}} 
	\left(\begin{array}{c}
		L\\
		k
	\end{array}\right)
	=\frac{1}{k!},
\end{eqnarray}
because every interaction vertex comes with a $1/L$ prefactor. If, however, there are more vertices in a diagram than there are iteration steps
to construct it (because we have connected several parts of $g_{l,L}$ both containing vertices by a new vertex), then the interactions still come with a prefactor
$1/L$, but we only have $\left(\begin{array}{c} L\\k^{\prime}\end{array}\right)$ possibilities to pick them, with $k^{\prime}<k$ and so these contributions
vanish in the limit of $L$ tending to infinity. Therefore, we can
neglect this possibility.\\

We remember that $g_{l,L}$ is a functional of $j$ and we obtain $\Gs^{\ii}$ only after inserting $j=\frac{\delta\Gs^{0}}{\delta\phi}$ (and taking the limit of $L \rightarrow \infty$). This becomes particularly important when discussing the effect of the lines \prettyref{eq:iteration_1pr_remove} and \prettyref{eq:iteration_2pr_remove}. They lead to diagrams consisting of one cluster stemming from the derivative of $\Gs$ with respect to $\phi$ or $\Delta$, which are attached to the remainder via one or two cumulants, respectively. The subtraction of $\phi$ in Eqs.~\prettyref{eq:iteration_1pr_remove} and \prettyref{eq:iteration_2pr_remove} removes all diagrams coming about by attaching the $\frac{\del \Gs}{\del \phi}$-cluster to isolated first-order cumulants of the unperturbed theory:\\ 
\begin{fmffile}{Removal_phi_deriv_Gamma_first_step}	
	\begin{eqnarray}
		\fl \mkern 40mu \parbox{50mm}{
			\begin{fmfgraph*}(70,30)
				\fmfpen{0.75thin}
				\fmfleft{i1}
				\fmfright{o1}
				\fmf{plain, tension=1.}{i1,o1}
				\fmfv{decor.shape=circle, decor.filled=empty, decor.size = thin}{o1}
				\fmfv{label=$\frac{\del \Gamma}{\del \phi}$, l.a=90, label.dist=10.}{i1}
				\fmfv{d.s=circle, d.filled=shaded}{i1}
				\fmfv{label=$\frac{\del W_{0}}{\del j}$, l.a=90, label.dist=10.}{o1}
			\end{fmfgraph*}		
		}
		\mkern-70mu
		-
		\mkern 30mu
		\parbox{50mm}{
			\begin{fmfgraph*}(70,30)
				\fmfpen{0.75thin}
				\fmfleft{i1}
				\fmfright{o1}
				\fmf{plain, tension=1.}{i1,o1}
				\fmfv{decor.shape=circle, decor.filled=full, decor.size = thin}{o1}
				\fmfv{label=$\frac{\del \Gamma}{\del \phi}$, l.a=90, label.dist=10.}{i1}
				\fmfv{d.s=circle, d.filled=shaded}{i1}
				\fmfv{label=$\phi$, l.a=90, label.dist=10.}{o1}
			\end{fmfgraph*}		
		}
		\mkern-80mu
		\overset{j=\frac{\del \Gamma_{0}}{\del \phi}}{=} 0 
		\label{eq:Cancelation_deriv_phi_Gamma}
	\end{eqnarray}
\end{fmffile}
Note that the cancelation only gets effective upon inserting $j=\frac{\delta\Gamma^{0}}{\delta\phi}$. We indicate this by denoting derivatives of $W_{0}$ evaluated at a variable $j$ by open circles and proper cumulants (the derivatives evaluated at $j=\frac{\delta\Gamma^{0}}{\delta\phi}$) by filled circles. This indicates that these two contributions are equal after inserting the proper value for $j$. However, the contribution from both diagrams in Eq.~\prettyref{eq:Cancelation_deriv_phi_Gamma} remain in $g_{l,L}$, but $j$-derivatives act only on the first one in further iteration steps.\\ 

The treatment of $j$-derivatives acting on $g_{l,L}$ for $l>0$, especially in line \prettyref{eq:iteration_2pr_remove} containing the $\Delta$ derivative is a bit more subtle.
We start by discussing its role in the Gaussian case, for which it is nearly analogous to that of line \prettyref{eq:iteration_1pr_remove}.

\section{Recovering the expansion around a Gaussian theory}\label{sec:Gaussian_theory_main}
Compared to Eq.~(13) in \cite{Kuehn18_375004}, we have obtained an additional contribution stemming from the two-point source. If the unperturbed theory is Gaussian, we can therefore proceed largely analogously to the first Legendre transform. Then,  Eq.~\prettyref{eq:iteration_1pr_remove} removes one-particle reducible diagrams and Eq.~\prettyref{eq:iteration_2pr_remove} removes two-particle reducible diagrams, as we will show in the following.\\

We have already seen in Eq.~\prettyref{eq:Cancelation_deriv_phi_Gamma} that, as a rule, Eq.~\prettyref{eq:iteration_1pr_remove} does not contribute in the first step of the iteration. However, the starting point for the iteration could still be ill-defined in case Eq.~\prettyref{eq:iteration_2pr_remove} already contributed at this stage. To avoid this, we have to make sure that also the second of these lines does not yield any nonzero contribution in case the $j$-derivatives solely act on $W_{0}$ and not on a diagram with interactions. 
To achieve this, the propagator of the unperturbed theory is chosen to be $\Delta$, so that, being applied to $W_{0}$ only, \prettyref{eq:iteration_2pr_remove} vanishes as well. Therefore, in the first iteration step, indeed only \prettyref{eq:iteration_generate_W} contributes to the iteration. In the next steps, the derivatives of $\Gs^{\ii}$ with respect to $\phi$ and $\Delta$ are easy to evaluate because $\Gs^{\ii}$ is a multinomial in $\phi$ and $\Delta$. Diagrammatically, a $\phi$-derivative then correponds to the removal of a dangling leg and a $\Delta$-derivative to the removal of a propagator line, {e.g.}

\begin{fmffile}{Derivatives_Gamma_diagrammatical}	
	\begin{eqnarray}
		\fl \frac{\del}{\del \phi}
		\mkern 10mu
		\parbox{25mm}{
			\begin{fmfgraph*}(50,25)
				\fmfpen{0.5thin}
				\fmfleft{i1}
				\fmfright{o1}
				\fmftop{t1}
				\fmfbottom{b1}
				\fmf{plain}{i1,v1,cm,v2,o1}
				\fmf{plain, tension=0., right=.5}{t1,v1,b1}
				\fmf{plain, tension=0., right=.5}{b1,v2,t1}
				\fmfv{decor.shape=circle,decor.filled=full, decor.size=6.5thin}{i1,b1,cm,t1,o1}

			\end{fmfgraph*}
		} 
		\mkern-20mu = 
		\parbox{25mm}{
			\begin{fmfgraph*}(50,25)
				\fmfpen{0.5thin}
				\fmfleft{i1}
				\fmfright{o1}
				\fmftop{t1}
				\fmfbottom{b1}
				\fmf{plain}{i1,v1,cm,v2,o1}
				\fmf{plain, tension=0., right=.5}{t1,v1,b1}
				\fmf{plain, tension=0., right=.5}{b1,v2,t1}
				\fmfv{decor.shape=circle,decor.filled=full, decor.size=6.5thin}{b1,cm,t1,o1}

			\end{fmfgraph*}
		} \mkern-15mu, \mkern 20mu
		\frac{\del}{\del \Delta}
		\mkern 10mu
		\parbox{25mm}{
			\begin{fmfgraph*}(50,25)
				\fmfpen{0.5thin}
				\fmfleft{i1}
				\fmfright{o1}
				\fmftop{t1}
				\fmfbottom{b1}
				\fmf{plain}{i1,v1,cm,v2,o1}
				\fmf{plain, tension=0., right=.5}{t1,v1,b1}
				\fmf{plain, tension=0., right=.5}{b1,v2,t1}
				\fmfv{decor.shape=circle,decor.filled=full, decor.size=6.5thin}{i1,b1,cm,t1,o1}
				
			\end{fmfgraph*}
		}
		\mkern-20mu =
		\mkern 10mu 
		\parbox{25mm}{
			\begin{fmfgraph*}(50,25)
				\fmfpen{0.5thin}
				\fmfleft{i1}
				\fmfright{o1}
				\fmftop{t1,e1,t2,e2,t3}
				\fmfbottom{b1}
				\fmf{plain}{i1,v1,cm,v2,o1}
				\fmf{plain, tension=0., right=.5}{v1,b1}
				\fmf{plain, tension=0., right=.5}{b1,v2}
				\fmf{plain, tension=0.}{v1,e1}
				\fmf{plain, tension=0.}{v2,e2}
				\fmfv{decor.shape=circle,decor.filled=full, decor.size=6.5thin}{i1,b1,cm,o1}
			
			\end{fmfgraph*}
		}
		\nonumber
	\end{eqnarray}
\end{fmffile}
(the additional prefactors emerging from the derivatives are implied in the symmetry factors of the diagrams on the right hand sides). Letting \prettyref{eq:iteration_1pr_remove} act on a cluster contained in some $g_{l^{\prime},L}$ for some $l^{\prime} \leq L$ generates a one-particle reducible diagram. The same is produced by \prettyref{eq:iteration_generate_W}, but with the opposite sign so that both contributions cancel out:
\begin{fmffile}{Contribution_phi_deriv_Gauss}	
	\begin{eqnarray}
		\fl \mkern 50mu \parbox{50mm}{
			\begin{fmfgraph*}(70,30)
				\fmfpen{0.75thin}
				\fmfleft{i1}
				\fmfright{o1}
				\fmf{plain, tension=1.}{i1,o1}
				\fmfv{decor.shape=circle, decor.filled=empty, decor.size = thin}{i1,o1}
				\fmfv{label=$g$, l.a=180, label.dist=10.}{i1}
				\fmfv{label=$g_{l^{\prime},,L}$, label.dist=10.}{o1}
			\end{fmfgraph*}		
		}
		\mkern -50mu
		- \mkern 50mu \parbox{50mm}{
			\begin{fmfgraph*}(70,30)
				\fmfpen{0.75thin}
				\fmfleft{i1}
				\fmfright{o1}
				\fmf{plain, tension=1.}{i1,o1}
				\fmfv{d.s=circle, d.filled=shaded}{i1}
				\fmfv{decor.shape=circle, decor.filled=empty, decor.size = thin}{o1}
				\fmfv{label=$\frac{\del \Gamma}{\del \phi}$, l.a=180, label.dist=10.}{i1}
				\fmfv{label=$g_{l^{\prime},,L}$, label.dist=10.}{o1}
			\end{fmfgraph*}		
		}
		\mkern -50mu
		\overset{j=\frac{\del \Gamma_{0}}{\del \phi}}{=} 0
		\label{eq:Cancellation_second_line_Gauss}
	\end{eqnarray}
\end{fmffile}
 
Line \prettyref{eq:iteration_2pr_remove} generates two kinds of contributions: the two $j$-derivatives can either act on the same cluster or on different ones. The first type of contributions (\prettyref{fig:Contribution_third_line_Gauss}, left) cancels two-particle reducible contributions emerging from line \prettyref{eq:iteration_generate_W}, whereas the second generates diagrams that are one-particle reducible at (at least) two points (\prettyref{fig:Contribution_third_line_Gauss}, right). The latter are again canceled by the corresponding diagrams emerging from line \prettyref{eq:iteration_1pr_remove}.\\
\begin{fmffile}{Contribution_Delta_deriv_Gauss}	
	\begin{figure}[h]
		\center
	        \parbox{50mm}{
			\begin{fmfgraph*}(70,30)
				\fmfpen{0.75thin}
				\fmfleft{i1}
				\fmfright{o1}
				\fmf{plain, tension=.25, right=.5}{i1,o1,i1}
				\fmfv{d.s=circle, d.filled=shaded}{i1}
				\fmfv{decor.shape=circle, decor.filled=empty, decor.size = thin}{o1}
				\fmfv{label=$\frac{\del \Gamma}{\del \Delta}$, label.dist=10.}{i1}
				\fmfv{label=$g_{l,,L}$, label.dist=10.}{o1}
			\end{fmfgraph*}		
		}
		\parbox{50mm}{
			\begin{fmfgraph*}(70,30)
				\fmfpen{0.75thin}
				\fmfleft{i1}
				\fmfright{o1}
				\fmf{plain, tension=1.}{i1,v1,o1}
				\fmfv{d.s=circle, d.filled=shaded}{v1}
				\fmfv{decor.shape=circle, decor.filled=empty, decor.size = thin}{i1,o1}
				\fmfv{label=$\frac{\del \Gamma}{\del \Delta}$, l.a=90, label.dist=10.}{v1}
				\fmfv{label=$g_{l,,L}$, label.dist=10.}{i1}
				\fmfv{label=$g_{l^{\prime},,L}$, label.dist=10.}{o1}
			\end{fmfgraph*}		
		}
		\caption{Types of contributions of \prettyref{eq:iteration_2pr_remove} to an interation step.}  
		\label{fig:Contribution_third_line_Gauss}
	\end{figure}
\end{fmffile}
In both cases of cancelation, we not only used that the diagrams are topologically identical, but also tacitly assumed that their symmetry factors were the same. Let us convince ourselves that this is actually the case. Consider a diagram with a cluster containing $k$ interactions connected to the remainder, containing $n-k$ interactions, by only one propagator line (or two, considering \prettyref{eq:iteration_2pr_remove}). In case we produce this diagram only using bare interactions, we have $\left(\begin{array}{c}n\\k\end{array}\right)$ possibilities to pick $k$ interactions for the cluster. This yields $\frac{1}{n!}\left(\begin{array}{c}n\\k\end{array}\right)=\frac{1}{k!\left(n-k\right)!}$ as a prefactor, same as when the cluster is contributed from the derivative of $\Gs$. The remaining symmetry operations fixing the symmetry prefactor are determined considering the cluster and the remainder separately. This indeed yields the same symmetry factor in either case to generate the diagram of the same topology.\\
Formally, we also could have chosen $\Gamma_{0}$ to be the second Legendre transform of the Gaussian theory and not its first. This approach is lengthier, but conceptually somewhat cleaner because by introducing the two-point source $K$ at this stage allows us to treat $\phi$ and $\Delta$ on the same footing: they are both determined by the choice of $j$ and $K$. In principle, this choice  also offers one way to generalize to correlated non-Gaussian theories, as we discuss in appendix \ref{General_non_Gauss}. This generalization, however, is only of limited interest because it requires the unperturbed theory to remain solvable after adding the two-point source. But if the unperturbed theory is uncorrelated, this is in general not the case. For the Ising model and a simple liquid, for example, the interaction has precisely the form of the two-point source term. We therefore have to treat this scenario separately, which we defer to \prettyref{sec:Uncorrelated_unperturbed_theory}.

\section{Uncorrelated unperturbed theory}\label{sec:Uncorrelated_unperturbed_theory}
If the unperturbed theory augmented with a two-point source term $K$ cannot exactly be solved, then we cannot cancel the corresponding term in the iteration equation, \prettyref{eq:iteration_2pr_remove} in the first step. This is not a problem, but certainly means  that the initial step of the iteration to generate $\Gs^{\ii}$ has to be different from the Gaussian case. What helps us here is that correlations are small and therefore, Eq.~\prettyref{eq:iteration_2pr_remove} can be treated order by order in the magnitude of the correlations. The only complication is that the concrete form of this perturbation is not {\it a priori} given, but it follows from the derivatives of the contribution to $\Gs^{\ii}$ from lower orders. For this reason, the zeroth, first and second order of $\Gs$ are special cases, so we treat them separately from higher order terms. We denote the $n$-th order contribution to $\Gs^{\ii}$ as $\Gs^{\ii,n}$.
\subsection{Zeroth, first and second order}
For simplicity, we  first assume only a single type of interaction, namely the two-point one. It can be absorbed into $\frac{\del\Gs}{\del \Delta}$ so that line \prettyref{eq:iteration_generate_W} is removed from the iteration equation:
\begin{eqnarray}
	\fl A\left[\frac{\del}{\del j}\right] \overset{\mathrm{two-point}\ \mathrm{interaction}}{=}  \frac{\del\Gs^{\ii}}{\del\phi}^{\T}\left(\frac{\del}{\del j}-\phi\right) + \frac{1}{2}\mathrm{tr}\frac{\del\Gs^{\ii}}{\del\Delta}\left[\left(\frac{\del}{\del j}-\phi\right)\left(\frac{\del}{\del j}-\phi\right)^{\T}-\Delta\right],
	\label{eq:Def_A_diff_op_two_point_int}
\end{eqnarray}
where, for this discussion of expansions around noninteracting theories, the trace operator excludes the diagonal term:
\begin{eqnarray}
	\mathrm{tr}\left(AB\right) = \int_{x\neq x^{\prime}} A_{x x^{\prime}}B_{x x^{\prime}}.
\end{eqnarray}
One can read off already from this form of the operator that diagrams containing "dangling legs", that is cumulants of order $1$, are canceled because every $j$-derivative is accompanied by a $-\phi$ term. 

\paragraph{Zeroth order} Evaluating Eq.~\prettyref{eq:Def_Gamma2V_implicit} at zeroth order in the interaction clearly yields $1$ on the left hand side. This tells us that the application of $A\left[\frac{\delta}{\delta j}\right]$ has to yield $0$ at this step. Because we reduce $\Gs^{\ii}$ by one order in the interaction by differentiating with respect to $\Delta$, we have 
\[
	\fl A\left[\frac{\delta}{\delta j}\right] = \frac{1}{2}\mathrm{tr}\frac{\del\Gs^{\ii,1}}{\del\Delta}^{\T}\left[\left(\frac{\del}{\del j}-\phi\right)\left(\frac{\del}{\del j}-\phi\right)^{\T}-\Delta\right] + \mathcal{O}\left(\Delta\right).
\] 
We conclude that the right-hand side of Eq.~\prettyref{eq:Def_Gamma2V_implicit} only vanishes to zeroth order in $\Delta$ if $\frac{\del\Gs^{\ii,1}}{\del\Delta} = 0$ and therefore $\Gs^{\ii,1}=0$ ($\Gs^{\ii,1}$ must vanish because it is proportional to $\Delta$ by definition). 
\paragraph{First order} Comparing the first order on both sides, we have $1$ from $\Gs^{\ii,1}=0$ on the left-hand side and
\begin{eqnarray}
	\fl &e^{-W_{0}\left[j\right]}\Big(1+\underbrace{\frac{\del\Gs^{\ii,1}}{\del\phi}^{\T}}_{=0}\left(\frac{\del}{\del j}-\phi\right)+\frac{1}{2}\mathrm{tr}\left(\frac{\del\Gs^{\ii,1}}{\del\Delta} + \frac{\del\Gs^{\ii,2}}{\del\Delta} \right) \left[\left(\frac{\del}{\del j}-\phi\right)\left(\frac{\del}{\del j}-\phi\right)^{\T}-\Delta\right]\Big) e^{W_{0}\left[j\right]} \nonumber\\
	\fl =&e^{-W_{0}\left[j\right]}\Big(1 + \frac{1}{2} \mathrm{tr}\frac{\del\Gs^{\ii,2}}{\del\Delta} \left(\frac{\del}{\del j}-\phi\right)\left(\frac{\del}{\del j}-\phi\right)^{\T}\Big) e^{W_{0}\left[j\right]} + \mathcal{O}\left(\Delta^2\right)
\end{eqnarray}
on the right-hand side. All nontrivial terms of order $\Delta$ vanish, as they should, because the $j$-derivatives of $W_{0}$ yield $\phi$ after evaluation (note that $\mathrm{tr}\frac{\del\Gs^{\ii,2}}{\del\Delta} \Delta$ is of order $2$). 
\paragraph{Second order} This is the first non-vanishing order. Its most general form reads
\begin{eqnarray}
	\fl \Gs^{\ii,2}\left[\phi,\Delta\right] = \frac{1}{2} \frac{1}{2^{2}} \int_{x\neq y} \int_{x^{\prime}\neq y^{\prime}} F_{xy;x^{\prime}y^{\prime}}\Delta_{xy}\Delta_{x^{\prime}y^{\prime}},\ F_{xy;x^{\prime}y^{\prime}} = F_{x^{\prime}y^{\prime};xy},\label{eq:Ansatz_GammaV2}
\end{eqnarray}
where we separated the $\frac{1}{2}$ prefactors from the definition of $F$ for later convenience. We observe that
\begin{eqnarray}
	\fl \frac{1}{2}\mathrm{tr}\Delta\frac{\del \Gs^{\ii,2}}{\del \Delta}\left[\phi,\Delta\right]  =  \frac{1}{2} \cdot \frac{1}{2} \int_{x\neq y}\int_{x^{\prime}\neq y^{\prime}} F_{xy;x^{\prime}y^{\prime}} \Delta_{xy}\Delta_{x^{\prime}y^{\prime}} = 2\, \Gs^{\ii,2}\left[\phi,\Delta\right]
\end{eqnarray}
To evaluate \prettyref{eq:Def_Gamma2V_implicit} neglecting terms of order $3$ or higher, we first rewrite its second-order expansion, bringing the last term in the operator \prettyref{eq:Def_A_diff_op}, -$\frac{1}{2}\mathrm{tr}\frac{\del \Gs^{\ii,2}}{\del \Delta}\left[\phi,\Delta\right]\Delta$, to the left-hand side so that we obtain (using again that $\Gs^{\ii,1}=0$)
\begin{eqnarray}
	\label{eq:Expansion_second_order_Gamma2V}
	\fl \exp\left(\Gamma^{\ii,2}\left[\phi,\Delta\right]\right)
	=  \exp\left(-W_{0}\left[j\right]\right) \exp\left(A_{2}\left[\frac{\del}{\del j}\right]\right) \exp\left(W_{0}\left[j\right]\right) + \mathcal{O}\left(\Delta^3\right),
\end{eqnarray}
where
\begin{eqnarray}
	\fl A_{2}\left[\frac{\del}{\del j}\right] :=  \frac{\del\Gs^{\ii,2}}{\del\phi}^{\T}\left(\frac{\del}{\del j}-\phi\right)
						    + \frac{1}{2}\mathrm{tr}\left(\frac{\del\Gs^{\ii,2}}{\del\Delta} + \frac{\del\Gs^{\ii,3}}{\del\Delta}\right) \left(\frac{\del}{\del j}-\phi\right)\left(\frac{\del}{\del j}-\phi\right)^{\T}.
	\label{eq:Def_A2_diff_op}
\end{eqnarray}
Here again, acting with the $A$-operator on $W_{0}$ just once yields zero, so that the only non-vanishing contribution is
\newpage
\begin{eqnarray}
	\fl \Gamma^{\ii,2} &= e^{-W_{0}\left[j\right]}\frac{1}{2} \int_{x}\int_{y} \frac{1}{2} \frac{\del\Gs^{\ii,2}}{\del\Delta_{xy}}  \left(\frac{\del}{\del j_{x}}-\phi_{x}\right)\left(\frac{\del}{\del j_{y}}-\phi_{y}\right)\\ 
	\fl &\int_{x^{\prime}}\int_{y^{\prime}}\frac{1}{2}\frac{\del\Gs^{\ii,2}}{\del\Delta_{x^{\prime}y^{\prime}}} \left(\frac{\del}{\del j_{x^{\prime}}}-\phi_{x^{\prime}}\right)\left(\frac{\del}{\del j_{y^{\prime}}}-\phi_{y^{\prime}}\right)e^{W_{0}\left[j\right]} \\
	\fl &=  \frac{1}{4}\int_{x\neq y} \left(\frac{\del\Gs^{\ii,2}}{\del\Delta_{xy}}\right)^{2}  \kv^{x} \kv^{y},\label{eq:Definition_Gamma2V_self_consistent}
\end{eqnarray}
where we introduced the second cumulants of the unperturbed theory, $\kv^{x}$. Correspondingly, we call the $n$-th order cumulants of the unperturbed theory $\kappa_{n}^{x}$. Diagrammatically, the right-hand side is represented by

\begin{fmffile}{TAP}	
	\begin{eqnarray}
        \fl \mkern-30mu \parbox{25mm}{
			\begin{fmfgraph*}(75,25)
				\fmfpen{0.5thin}
				\fmftop{o1,o2,o3,o4,o5}
				\fmfbottom{u1,u2,u3,u4,u5}
				\fmf{phantom}{u1,v1,o3}
				\fmf{plain}{v1,o3}
				\fmf{phantom}{o1,v1,u3}
				\fmf{plain}{v1,u3}
				\fmf{phantom}{u3,v2,o5}
				\fmf{plain}{u3,v2}
				\fmf{phantom}{o3,v2,u5}
				\fmf{plain}{o3,v2}
				\fmfv{decor.shape=circle,decor.filled=empty, decor.size=6.5thin}{v1,v2}
			\end{fmfgraph*}
			}
		& \mkern-20mu = \frac{1}{2!2^{2}} 2 \int_{x\neq y} \left(\frac{\del\Gs^{\ii,2}}{\del\Delta_{xy}}\right)^{2} \kv^{x} \kv^{y},  \label{eq:Def_second_oder_diagram}
	\end{eqnarray}
\end{fmffile}
where we introduced the notation
\begin{fmffile}{Definition_edge_second_Legendre}	
	\begin{eqnarray}
		\fl \J_{xy} := \frac{\del}{ \del \Delta_{xy}} \Gs^{\ii,2} := \parbox{25mm}{
			\begin{fmfgraph*}(25,25)
				\fmfpen{0.5thin}
				\fmftop{o1,o2,o3}
				\fmfbottom{u1,u2,u3}
				\fmf{plain}{u1,o2}
				\fmf{plain}{u3,o2}
				\fmfv{label=$x$, label.angle=-90, label.dist=4.5pt}{u1}
				\fmfv{label=$y$, label.angle=-90, label.dist=4.5pt}{u3}
			\end{fmfgraph*}
		}
		\label{eq:Def_diagram_simple_interation} 	
	\end{eqnarray}
\end{fmffile}
and the prefactors on the right-hand side of Eq.~\prettyref{eq:Def_second_oder_diagram} indicate how we evaluate the diagram. The $\frac{1}{2}$ comes from the order in the interaction, the $\frac{1}{2^{2}}$ from the definition of the interaction and the $2$ is the symmetry-factor of the diagram: flipping one interaction vertex generates a new labeled diagram, flipping both at once does not. Inserting the form Eq.~\prettyref{eq:Ansatz_GammaV2} into Eq.~\prettyref{eq:Definition_Gamma2V_self_consistent}, we obtain the tensor equation
\begin{eqnarray}
	\fl F_{x^{\prime}y^{\prime};x^{\prime\prime}y^{\prime\prime}} = \frac{1}{2} \int_{x\neq y} F_{xy;x^{\prime}y^{\prime}}F_{xy;x^{\prime\prime} y^{\prime\prime}} \kv^{x}\kv^{y},
\end{eqnarray}
which is solved by
\begin{eqnarray}
	\fl F_{xy;x^{\prime}y^{\prime}} =  2\delta_{xx^{\prime}}\delta_{yy^{\prime}}\left(\kv\right)^{-1}_{x} \left(\kv\right)^{-1}_{y}
\end{eqnarray}
and, consequently,
\begin{eqnarray}
	\fl \Gs^{\ii,2} &= \frac{1}{4}\int_{x\neq y}\Delta_{xy}^{2}  \left(\kv\right)^{-1}_{x} \left(\kv\right)^{-1}_{y} \label{eq:Second_order_correction}\\
	\fl \J_{xy} &= \left(\kv\right)^{-1}_{x}\Delta_{xy}\left(\kv\right)^{-1}_{y} .\label{eq:Def_coupling}
\end{eqnarray}
In higher orders, $\J$ will play the same role in the diagrammatics as the two-point interaction does for the first Legendre transform.

\subsection{Third and higher orders}\label{subsec:terramystica}
Before proceeding to constructing further diagrams, we recall the meaning of their constituting elements defined so far and introduce two new ones:
\vspace*{2mm}
\begin{fmffile}{Collection_diagram_elements}	
	\begin{eqnarray}
		\fl
		\parbox[35mm]{25mm}{
			\begin{fmfgraph*}(25,25)
				\fmfpen{0.5thin}
				\fmftop{o1,o2,o3,o4,o5}
				\fmfbottom{u1,u2,u3,u4,u5}
				\fmf{phantom}{u1,m1,o1}
				\fmf{phantom}{u3,m3,o3}
				\fmf{phantom}{u5,m5,o5}
				\fmf{plain}{m3,o4}
				\fmf{plain}{m3,o5}
				\fmf{plain,tension=0}{m3,m5}
				\fmf{phantom, tension=10}{m1,m3}
				\fmf{plain,tension=2.5}{m3,u5}
				\fmfv{decor.shape=circle,decor.filled=empty, decor.size=6.5thin}{m3}
				\fmfv{label=$x$, label.angle=90, label.dist=4.5pt}{o4}
				\fmfv{label=$y$, label.angle=60, label.dist=2.pt}{o5}
				\fmfv{label=$z$, label.angle=20, label.dist=1.5pt}{m5}
				\fmfv{label=$...$, label.angle=0, label.dist=1.5pt}{u5}
			\end{fmfgraph*}
		}  \mkern-55mu = \frac{\delta^{n} W}{\delta j_{x}\delta j_{y}\delta j_{z}...}, 
		\parbox[35mm]{25mm}{
			\begin{fmfgraph*}(25,25)
				\fmfpen{0.5thin}
				\fmftop{o1,o2,o3,o4,o5}
				\fmfbottom{u1,u2,u3,u4,u5}
				\fmf{phantom}{u1,m1,o1}
				\fmf{phantom}{u3,m3,o3}
				\fmf{phantom}{u5,m5,o5}
				\fmf{plain}{m3,o4}
				\fmf{plain}{m3,o5}
				\fmf{plain,tension=0}{m3,m5}
				\fmf{phantom, tension=10}{m1,m3}
				\fmf{plain,tension=2.5}{m3,u5}
				\fmfv{decor.shape=circle,decor.filled=full, decor.size=6.5thin}{m3}
				\fmfv{label=$x$, label.angle=90, label.dist=4.5pt}{o4}
				\fmfv{label=$y$, label.angle=60, label.dist=2.pt}{o5}
				\fmfv{label=$z$, label.angle=20, label.dist=1.5pt}{m5}
				\fmfv{label=$...$, label.angle=0, label.dist=1.5pt}{u5}
			\end{fmfgraph*}
		}  \mkern-55mu = \kappa^{n}_{x,y,z,...},
		\mkern 10mu 
		\parbox{7mm}{
			\begin{fmfgraph*}(15,15)
				\fmfpen{0.5thin}
				\fmfleft{l}
				\fmfright{r}
				\fmf{wiggly}{l,a,r}
				\fmfv{label=$x$, label.angle=-90, label.dist=4.5pt}{l}
				\fmfv{label=$x$, label.angle=-90, label.dist=4.5pt}{r}
				\fmfv{decor.shape=circle,decor.filled=shaded, decor.size=6.5thin}{a}
			\end{fmfgraph*}
		}:= \frac{\delta^{2}\Gamma_{0}}{\delta \phi_{x}^{2}}= \left(\kv\right)^{-1},
		\mkern 15mu
		\parbox{25mm}{
			\begin{fmfgraph*}(25,25)
				\fmfpen{0.5thin}
				\fmftop{o1,o2,o3}
				\fmfbottom{u1,u2,u3}
				\fmf{plain}{u1,o2}
				\fmf{plain}{u3,o2}
				\fmfv{label=$x$, label.angle=-90, label.dist=4.5pt}{u1}
				\fmfv{label=$y$, label.angle=-90, label.dist=4.5pt}{u3}
			\end{fmfgraph*}
		} \mkern-60mu = \J_{xy}. 
	\end{eqnarray}
\end{fmffile}
We recall that the difference between the empty and the full nodes is that the first represent derivatives of the cumulant-generating functional at arbitrary sources, whereas the full ones indicate that we have inserted $j = \frac{\partial \Gamma_{0}}{\partial \phi}$ - so these nodes represent the actual cumulants. Because the (first) Legendre transform is involutive, we furthermore have 
\begin{fmffile}{Relation_cumulant_anticumulant}
$\parbox{10mm}{
	\begin{fmfgraph*}(30,15)
		\fmfpen{0.5thin}
		\fmfleft{l}
		\fmfright{r}
		\fmf{phantom}{l,c,m,a,r}
		\fmf{plain}{l,c,m}
		\fmf{wiggly}{m,a,r}
		\fmfv{decor.shape=circle,decor.filled=empty, decor.size=6.5thin}{c}
		\fmfv{decor.shape=circle,decor.filled=shaded, decor.size=6.5thin}{a}
		\end{fmfgraph*}
}=1$
\end{fmffile}.\\
Singling out the terms stemming from the second order of $\Gs^{\ii}$ (using $\J_{xy} = \frac{\delta \Gamma^{\ii,2}}{\delta \Delta_{xy}}$), which we subtract and then add in a different form, we now start from
\begin{eqnarray}
 	\fl & \exp\left(-\Gamma^{\ii}\left[\phi,\Delta\right]\right)\nonumber \\
	\fl = & \exp\left(-W_{0}\left[j\right]\right)\exp\left(\int_{x} \frac{\del\Gamma^{\ii}}{\del\phi_{x}} \left(\frac{\del}{\del j_{x}}-\phi_{x}\right) + \frac{1}{2}\int_{x\neq y}\left(\frac{\del\left(\Gamma^{\ii}-\Gamma^{\ii,2}\right)}{\del\Delta_{xy}} + \J_{xy}\right)\right.\label{eq:Perturbing_part_Ising_second_Legendre_initial_form-1}\\
 	\fl & \left.\left.\left[\left(\frac{\del}{\del j_{x}}-\phi_{x}\right)\left(\frac{\del}{\del j_{y}}-\phi_{y}\right)-\Delta_{xy}\right]\right)\exp\left(W_{0}\left[j\right]\right)\right|_{j=j_{0}}.
\end{eqnarray}
which leads to the following iteration equation for $g_{l,L}$:
\begin{eqnarray}
	 \fl & g_{l+1,L}-g_{l,L}=\nonumber \\
 	 \fl   &\frac{1}{L}\,e^{-g_{l,L}\left[j,K\right]}\, \int_{x}\frac{\del\Gamma^{\ii}}{\del\phi_{x}} \left(\frac{\del}{\del j_{x}}-\phi_{x}\right) \,e^{g_{l,L}\left[j,K\right]}\label{eq:iteration_Gamma1_2p_interaction} \\
 	 \fl  + &\frac{1}{L}\,e^{-g_{l,L}\left[j,K\right]}\, \frac{1}{2}\int_{x\neq y}\left(\frac{\del\Gamma^{\ii,\neq 2}}{\del\Delta_{xy}} + \J_{xy}\right) \left[\left(\frac{\del}{\del j_{x}}-\phi_{x}\right)\left(\frac{\del}{\del j_{y}}-\phi_{y}\right)-\Delta_{xy}\right]\,e^{g_{l,L}\left[j,K\right]}\label{eq:iteration_Gamma2_2p_interaction} \\
 	 \fl  + &\mathcal{O}(L^{-2}),\label{eq:Iteration_g-concrete}
\end{eqnarray}
where we abbreviated $\Gs^{\ii,\neq 2} = \Gs^{\ii} - \Gs^{\ii,2}$. Calling $\Gs^{\ii,n}$ the interaction term of order $n$, the term $\frac{\del \Gs^{\ii,n}}{\del \Delta}$ is of order $n-1$, which means that, combined with $\J$ or the term $\Delta$ in Eq.~\prettyref{eq:iteration_Gamma2_2p_interaction}, it yields a term of order $n$. However, we see that these two contributions actually cancel out because
\begin{eqnarray}
	\fl &e^{-W_{0}\left[j\right]}\frac{1}{2}\int_{x\neq y}\frac{\del\Gamma^{\ii,n}}{\del\Delta_{xy}} \left(\frac{\del}{\del j_{x}}-\phi_{x}\right)\left(\frac{\del}{\del j_{y}}-\phi_{y}\right) \frac{1}{2}\int_{x^{\prime}\neq y^{\prime}}\J_{x^{\prime}y^{\prime}} \left(\frac{\del}{\del j_{x^{\prime}}}-\phi_{x^{\prime}}\right)\left(\frac{\del}{\del j_{y^{\prime}}}-\phi_{y^{\prime}}\right)\,e^{W_{0}\left[j\right]}\nonumber\\
	\fl=& \frac{1}{2}\int_{x\neq y}\frac{\del\Gamma^{\ii,n}}{\del\Delta_{xy}} \Delta_{xy}.\label{eq:Implicit_contribution_canceling_term}
\end{eqnarray}
This result comes about by observing that any $W_{0}$-derivative generated in Eq.~\prettyref{eq:Implicit_contribution_canceling_term} of order higher than $2$ is multiplied by at least one first-order one, so those terms are eventually canceled by inserting $j=\frac{\delta\Gamma}{\delta \phi}$ and considering the respective $(-\phi)$-contribution. Therefore, the only non-vanishing contribution arises from terms in which either $x=x^{\prime},\,y=y^{\prime}$ or $x=y^{\prime},\,y=x^{\prime}$ and the respective pairs of $j$-derivatives generate two second-order cumulants. Due to these two possibilities, one $\frac{1}{2}$ is turned into unity. We therefore obtain the contributions to the $n$-th order by combining clusters from $\frac{\del \Gs^{k}}{\del \Delta}$ for $k<n$ in all possible ways such that the total number of correlations in the diagram equals $n$. This enables the explicit iterative construction of all higher-order contributions, which is one of the main results of this work. 

\paragraph{Third order} Here, concretely, we construct all diagrams without dangling legs containing $\J_{xy}$ three times. This yields the same diagrams as for the first Legendre transform. We  explicitly write the minus sign appearing in the left-hand side of Eq.~\prettyref{eq:Gamma_from_g}:
\begin{fmffile}{Third_order_Leg2}
	\begin{eqnarray}
		\fl -\mkern-20mu \parbox{25mm}{
			\begin{fmfgraph*}(75,25)
				\fmfpen{0.5thin}
				\fmftop{o1,o2,o3}
				\fmfbottom{u1,u2,u3}
				\fmf{phantom,tension=100}{u1,dl,v1,o2}
				\fmf{plain}{dul,v1,o2}
				\fmf{phantom,tension=100}{u3,dr,v2,o2}
				\fmf{plain}{dur,v2,o2}
				\fmf{phantom}{u1,dul,u2,dur,u3}
				\fmf{plain}{dul,u2,dur}
				\fmf{phantom,tension=0.5}{dl,dum,dr}
				\fmf{phantom,tension=1}{dum,u2}
				\fmfv{decor.shape=circle,decor.filled=empty, decor.size=6.5thin}{v1,v2,u2}
			\end{fmfgraph*}
		}		
		& \mkern-60mu = -\frac{1}{3!2^{3}} 2^{3} \int_{x\neq y\neq z\neq x} \J_{xy} \J_{yz} \J_{zx}\kv^{x}\kv^{y}\kv^{z}\label{eq:Def_third_order_ring_diagram}&\\
		\fl -\mkern 20mu \parbox{25mm}{
			\begin{fmfgraph*}(25,25)
				\fmfpen{0.5thin}
				\fmftop{o1,o2,o3}
				\fmfbottom{u1,u2,u3}
				\fmf{phantom}{u2,du,do,o2}
				\fmf{phantom,tension=2}{du,do,o2}
				\fmf{plain}{u1,o2}
				\fmf{plain}{u3,o2}
				\fmf{plain}{u1,do}
				\fmf{plain}{u3,do}
				\fmf{plain}{u1,du}
				\fmf{plain}{u3,du}
				\fmfv{decor.shape=circle,decor.filled=empty, decor.size=6.5thin}{u1,u3}
			\end{fmfgraph*}
		}
		& \mkern-60mu = -\frac{1}{3!2^{3}} 2^{2} \int_{x\neq y} \J_{xy}^{3} \ks^{x}\ks^{y}.\label{eq:Def_third_order_Citroen_diagram}&
	\end{eqnarray}
\end{fmffile}
These diagrams have the "ring" and the  "watermelon" topology also encountered in Feynman diagrammatics.
For the construction of higher orders, we have to consider $\Delta$-derivatives of the third-order contributions to $\Gs^{\ii}$ as additional interactions. We therefore want to determine the symmetry factor of clusters with one $\Delta$ correlation removed. Since we perform a $\Delta$ derivative, we have to fix a pair of outer indices. Because correlations are symmetric, this yields a factor $2$, which is compensated by the $\frac 12$ factor  in Eq.~\prettyref{eq:iteration_2pr_remove}. In practice, this means, for example, that flipping both interactions at once in the diagram of Eq.~\prettyref{eq:Diagram_cherry_ring_third_order} with two "outer legs" generates a new labeled diagram. Concretely, we translate
\begin{fmffile}{Deriv_Third_order_Leg2}
	\begin{eqnarray}
		\fl \frac{\del}{\del \Delta_{xy}} (-\mkern-20mu \parbox{25mm}{
			\begin{fmfgraph*}(75,25)
				\fmfpen{0.5thin}
				\fmftop{o1,o2,o3}
				\fmfbottom{u1,u2,u3}
				\fmf{phantom,tension=100}{u1,dl,v1,o2}
				\fmf{plain}{dul,v1,o2}
				\fmf{phantom,tension=100}{u3,dr,v2,o2}
				\fmf{plain}{dur,v2,o2}
				\fmf{phantom}{u1,dul,u2,dur,u3}
				\fmf{plain}{dul,u2,dur}
				\fmf{phantom,tension=0.5}{dl,dum,dr}
				\fmf{phantom,tension=1}{dum,u2}
				\fmfv{decor.shape=circle,decor.filled=full, decor.size=6.5thin}{v1,v2,u2}
			\end{fmfgraph*}
		}
		\mkern-10mu)
		=
		-\mkern 10mu
		\parbox{20mm}{
			\begin{fmfgraph*}(20,30) 				
				\fmfpen{0.5thin} 				
				\fmfleft{l1,l2,l3,l4,l5} 				
				\fmfright{r1,r2,c1,r4,r5}
				\fmf{plain}{r1,l2}
				\fmf{plain}{l2,c1}
				\fmf{plain}{c1,l4}
				\fmf{plain}{l4,r5}				
				\fmfv{decor.shape=circle,decor.filled=full, decor.size=6.5thin}{c1} 			
			\end{fmfgraph*} 		
		}
		\mkern-40mu
		=
		-\int_{z\left(\neq x,y\right)}\J_{xz}\kv^{z}\J_{zy}, \label{eq:Diagram_deriv_ring_third_order}\\
		\fl \frac{\del}{\del \Delta_{xy}} (-\mkern 15mu
		\parbox{25mm}{
			\begin{fmfgraph*}(25,25)
				\fmfpen{0.5thin}
				\fmftop{o1,o2,o3}
				\fmfbottom{u1,u2,u3}
				\fmf{phantom}{u2,du,do,o2}
				\fmf{phantom,tension=2}{du,do,o2}
				\fmf{plain}{u1,o2}
				\fmf{plain}{u3,o2}
				\fmf{plain}{u1,do}
				\fmf{plain}{u3,do}
				\fmf{plain}{u1,du}
				\fmf{plain}{u3,du}
				\fmfv{decor.shape=circle,decor.filled=full, decor.size=6.5thin}{u1,u3}
			\end{fmfgraph*}
		}
		\mkern-60mu)
		=
		-\mkern 10mu
		\parbox{25mm}{ 			
			\begin{fmfgraph*}(50,25) 				
				\fmfpen{0.5thin} 				
				\fmfleft{l1,l2,l3} 				
				\fmftop{t1,t2,c1,a1,t3}
				\fmfbottom{b1,b2,c2,a2,b3}
				\fmf{phantom}{t2,do,b2}
				\fmf{phantom}{do,l2}
				\fmf{plain}{c1,l2} 				
				\fmf{plain}{c2,l2} 				
				\fmf{plain}{c1,do} 				
				\fmf{plain}{c2,do} 				
				\fmf{wiggly}{t3,a1,c1} 				
				\fmf{wiggly}{b3,a2,c2} 				
				\fmfv{decor.shape=circle,decor.filled=shaded, decor.size=6.5thin}{a1,a2} 				
				\fmfv{decor.shape=circle,decor.filled=full, decor.size=6.5thin}{c1,c2} 			
			\end{fmfgraph*} 		
		}
		\mkern -20mu
		=
		-\frac{1}{2}\frac{\ks^{x}}{\kv^{x}}\J_{xy}^{2}\frac{\ks^{y}}{\kv^{y}}.\label{eq:Diagram_cherry_ring_third_order}	
	\end{eqnarray}
\end{fmffile}
We call subdiagrams of the types presented in Eq.~\prettyref{eq:Diagram_cherry_ring_third_order} "caterpillar" and "watermelon".
Note that the sign of the contributions changes from the second to the third order - this is different in the first Legendre transform, where also the second order contributes with a negative sign.

\paragraph{Fourth order} One possibility to construct a fourth-order diagram using these components is to take one sub-diagram of each of the two building blocks in Eq.~\prettyref{eq:Diagram_cherry_ring_third_order}. Because we are free also to rebuild each sub-diagram using first-order interactions, defined in Eq.~\prettyref{eq:Def_diagram_simple_interation}, we have to sum up four contributions in total:
\begin{fmffile}{Terra_mystica_diagrams} 	
	\begin{eqnarray}		 		
		&-\parbox{25mm}{ 			
			\begin{fmfgraph*}(100,25) 				
				\fmfpen{0.5thin} 				
				\fmfleft{l1,l2,l3,l4,l5}
				\fmfright{r1,r2,r3}				
				\fmftop{t1,t2,c1,a1,cv,aa1,cc1,t3,t4}
				\fmfbottom{b1,b2,c2,a2,ccv,aa2,cc2,b3,b4}
				\fmf{phantom}{t2,dol,b2}
				\fmf{phantom}{dol,r2}
				\fmf{phantom}{t3,dor,b3}
				\fmf{phantom}{dor,r2}
				\fmf{plain}{c1,l4} 				
				\fmf{plain}{l4,dol} 				
				\fmf{plain}{l2,dol} 				
				\fmf{plain}{c2,l2}
				\fmf{plain}{cc1,r2}
				\fmf{plain}{cc2,r2}
				\fmf{plain}{cc1,dor}
				\fmf{plain}{cc2,dor}				
				\fmf{wiggly}{c1,a1,cv,aa1,cc1} 				
				\fmf{wiggly}{c2,a2,ccv,aa2,cc2}
				\fmfv{decor.shape=circle,decor.filled=shaded, decor.size=6.5thin}{a1,a2,aa1,aa2} 				
				\fmfv{decor.shape=circle,decor.filled=full, decor.size=6.5thin}{dol,c1,c2,cc1,cc2}
				\fmfv{decor.shape=circle,decor.filled=empty, decor.size=6.5thin}{cv,ccv}			
			\end{fmfgraph*} 		
		}
		\mkern 58mu
		-
		\parbox{25mm}{ 			
			\begin{fmfgraph*}(50,25) 				
				\fmfpen{0.5thin} 				
				\fmfleft{l1,l2,l3,l4,l5}
				\fmfright{r1,r2,r3}				
				\fmftop{t1,t2,cc1,t3,t4}
				\fmfbottom{b1,b2,cc2,b3,b4}
				\fmf{phantom}{t2,dol,b2}
				\fmf{phantom}{dol,r2}
				\fmf{phantom}{t3,dor,b3}
				\fmf{phantom}{dor,r2}
				\fmf{plain}{cc1,l4} 				
				\fmf{plain}{l4,dol} 				
				\fmf{plain}{l2,dol} 				
				\fmf{plain}{cc2,l2}
				\fmf{plain}{cc1,r2}
				\fmf{plain}{cc2,r2}
				\fmf{plain}{cc1,dor}
				\fmf{plain}{cc2,dor}				
				\fmfv{decor.shape=circle,decor.filled=empty, decor.size=6.5thin}{cc1,cc2,dol}
			\end{fmfgraph*} 		
		}\\[5mm]
		+
		&\parbox{25mm}{ 			
			\begin{fmfgraph*}(100,25) 				
				\fmfpen{0.5thin} 				
				\fmfleft{l1,l2,l3,l4,l5}
				\fmfright{r1,r2,r3}				
				\fmftop{t1,t2,cv,aa1,cc1,t3,t4}
				\fmfbottom{b1,b2,ccv,aa2,cc2,b3,b4}
				\fmf{phantom}{t2,dol,b2}
				\fmf{phantom}{dol,r2}
				\fmf{phantom}{t3,dor,b3}
				\fmf{phantom}{dor,r2}
				\fmf{plain}{cv,l4} 				
				\fmf{plain}{l4,dol} 				
				\fmf{plain}{l2,dol} 				
				\fmf{plain}{ccv,l2}
				\fmf{plain}{cc1,r2}
				\fmf{plain}{cc2,r2}
				\fmf{plain}{cc1,dor}
				\fmf{plain}{cc2,dor}				
				\fmf{wiggly}{cv,aa1,cc1} 				
				\fmf{wiggly}{ccv,aa2,cc2}
				\fmfv{decor.shape=circle,decor.filled=shaded, decor.size=6.5thin}{aa1,aa2} 				
				\fmfv{decor.shape=circle,decor.filled=full, decor.size=6.5thin}{cc1,cc2}
				\fmfv{decor.shape=circle,decor.filled=empty, decor.size=6.5thin}{dol,cv,ccv}			
			\end{fmfgraph*} 		
		}
		\mkern 60mu
		+
			\parbox{25mm}{ 			
			\begin{fmfgraph*}(100,25) 				
				\fmfpen{0.5thin} 				
				\fmfleft{l1,l2,l3,l4,l5}
				\fmfright{r1,r2,r3}				
				\fmftop{t1,t2,c1,a1,cc1,t3,t4}
				\fmfbottom{b1,b2,c2,a2,cc2,b3,b4}
				\fmf{phantom}{t2,dol,b2}
				\fmf{phantom}{dol,r2}
				\fmf{phantom}{t3,dor,b3}
				\fmf{phantom}{dor,r2}
				\fmf{plain}{c1,l4} 				
				\fmf{plain}{l4,dol} 				
				\fmf{plain}{l2,dol} 				
				\fmf{plain}{c2,l2}
				\fmf{plain}{cc1,r2}
				\fmf{plain}{cc2,r2}
				\fmf{plain}{cc1,dor}
				\fmf{plain}{cc2,dor}
				\fmf{wiggly}{c1,a1,cc1}
				\fmf{wiggly}{c2,a2,cc2}
				\fmfv{decor.shape=circle,decor.filled=shaded, decor.size=6.5thin}{a1,a2}					
				\fmfv{decor.shape=circle,decor.filled=empty, decor.size=6.5thin}{cc1,cc2}
				\fmfv{decor.shape=circle,decor.filled=full, decor.size=6.5thin}{c1,c2,dol}			
			\end{fmfgraph*} 		
		}
		\mkern 60mu .	
	\end{eqnarray} 
\end{fmffile}
We evaluate these diagrams using Eqs.~\prettyref{eq:Def_diagram_simple_interation}, \prettyref{eq:Diagram_deriv_ring_third_order} and \prettyref{eq:Diagram_cherry_ring_third_order}, which yields for the second diagram
\begin{fmffile}{First_Terra_mystica_diagrams_evaluation} 	
	\begin{eqnarray}
		\fl -\parbox{25mm}{ 			
			\begin{fmfgraph*}(50,25) 				
				\fmfpen{0.5thin} 				
				\fmfleft{l1,l2,l3,l4,l5}
				\fmfright{r1,r2,r3}				
				\fmftop{t1,t2,cc1,t3,t4}
				\fmfbottom{b1,b2,cc2,b3,b4}
				\fmf{phantom}{t2,dol,b2}
				\fmf{phantom}{dol,r2}
				\fmf{phantom}{t3,dor,b3}
				\fmf{phantom}{dor,r2}
				\fmf{plain}{cc1,l4} 				
				\fmf{plain}{l4,dol} 				
				\fmf{plain}{l2,dol} 				
				\fmf{plain}{cc2,l2}
				\fmf{plain}{cc1,r2}
				\fmf{plain}{cc2,r2}
				\fmf{plain}{cc1,dor}
				\fmf{plain}{cc2,dor}				
				\fmfv{decor.shape=circle,decor.filled=empty, decor.size=6.5thin}{cc1,cc2,dol}
			\end{fmfgraph*} 		
		}  & \mkern -25mu = -\frac{1}{4!} \left(\begin{array}{c}4 \nonumber\\
2
\end{array}\right) \cdot \left[ 2^{2}\cdot \left(\frac{1}{2}\right)^{2}\right] \cdot \left[ 2 \cdot \left(\frac{1}{2}\right)^{2}\right] \cdot 2 \int_{x\neq y\neq z\neq x}\ks^{x}\J_{xy}^{2}\ks^{y}\J_{xz}\kv^{z}\J_{zy} \\[5mm]
		& \mkern -25mu = -\frac{1}{4} \int_{x\neq y\neq z\neq x}\ks^{x}\J_{xy}^{2}\ks^{y}\J_{xz}\kv^{z}\J_{zy}\label{eq:First_TM_diagram},
	\end{eqnarray} 
\end{fmffile}
where we made the origin of the prefactor fully explicit. Because there are four interactions, there is a  $\frac{1}{4!}\left(\frac{1}{2}\right)^{4}$ prefactor. Then, we have $\left(\begin{array}{c}4\\
2
\end{array}\right)$ possibilities to choose two vertices for one of the subdiagrams. In the caterpillar-diagram, we can switch both interactions to generate a new labeled one, whereas in the other one, we can only flip one of them. Finally, there are two possibilities to join both subdiagrams.\\

Evaluating the diagrams joining two second-order contributions, we obtain
\begin{fmffile}{Second_Terra_mystica_diagrams_evaluation} 	
	\begin{eqnarray}		 		
		&-\parbox{25mm}{ 			
			\begin{fmfgraph*}(100,25) 				
				\fmfpen{0.5thin} 				
				\fmfleft{l1,l2,l3,l4,l5}
				\fmfright{r1,r2,r3}				
				\fmftop{t1,t2,c1,a1,cv,aa1,cc1,t3,t4}
				\fmfbottom{b1,b2,c2,a2,ccv,aa2,cc2,b3,b4}
				\fmf{phantom}{t2,dol,b2}
				\fmf{phantom}{dol,r2}
				\fmf{phantom}{t3,dor,b3}
				\fmf{phantom}{dor,r2}
				\fmf{plain}{c1,l4} 				
				\fmf{plain}{l4,dol} 				
				\fmf{plain}{l2,dol} 				
				\fmf{plain}{c2,l2}
				\fmf{plain}{cc1,r2}
				\fmf{plain}{cc2,r2}
				\fmf{plain}{cc1,dor}
				\fmf{plain}{cc2,dor}				
				\fmf{wiggly}{c1,a1,cv,aa1,cc1} 				
				\fmf{wiggly}{c2,a2,ccv,aa2,cc2}
				\fmfv{decor.shape=circle,decor.filled=shaded, decor.size=6.5thin}{a1,a2,aa1,aa2} 				
				\fmfv{decor.shape=circle,decor.filled=full, decor.size=6.5thin}{dol,c1,c2,cc1,cc2}
				\fmfv{decor.shape=circle,decor.filled=empty, decor.size=6.5thin}{cv,ccv}			
			\end{fmfgraph*} 		
		}
		\mkern 58mu = -\frac{1}{4}\int_{x\neq y\neq z\neq x}\ks^{x}\J_{xy}^{2}\ks^{y}\J_{xz}\kv^{z}\J_{zy}, 
	\end{eqnarray} 
\end{fmffile}
where the prefactor is explained as follows: The two entire subdiagrams of order $2$ act as interactions so they come with a $\frac{1}{2}$. Analogously to the respective subdiagram in Eq.~\prettyref{eq:First_TM_diagram} (corresponding to the second square bracket), the watermelon contribution comes with a $\frac{1}{2}$ and we have two possibilites to join the legs of the two interactions. Together this leads to a $2\cdot \left(\frac{1}{2}\right)^{2}\frac{1}{2}=\frac{1}{4}$ weight.\\

Along similar considerations, we evaluate the two remaining contributions, which yield
\begin{fmffile}{Further_Terra_mystica_diagrams_evaluation} 	
	\begin{eqnarray}
		\fl &\parbox{25mm}{ 			
			\begin{fmfgraph*}(100,25) 				
				\fmfpen{0.5thin} 				
				\fmfleft{l1,l2,l3,l4,l5}
				\fmfright{r1,r2,r3}				
				\fmftop{t1,t2,cv,aa1,cc1,t3,t4}
				\fmfbottom{b1,b2,ccv,aa2,cc2,b3,b4}
				\fmf{phantom}{t2,dol,b2}
				\fmf{phantom}{dol,r2}
				\fmf{phantom}{t3,dor,b3}
				\fmf{phantom}{dor,r2}
				\fmf{plain}{cv,l4} 				
				\fmf{plain}{l4,dol} 				
				\fmf{plain}{l2,dol} 				
				\fmf{plain}{ccv,l2}
				\fmf{plain}{cc1,r2}
				\fmf{plain}{cc2,r2}
				\fmf{plain}{cc1,dor}
				\fmf{plain}{cc2,dor}				
				\fmf{wiggly}{cv,aa1,cc1} 				
				\fmf{wiggly}{ccv,aa2,cc2}
				\fmfv{decor.shape=circle,decor.filled=shaded, decor.size=6.5thin}{aa1,aa2} 				
				\fmfv{decor.shape=circle,decor.filled=full, decor.size=6.5thin}{cc1,cc2}
				\fmfv{decor.shape=circle,decor.filled=empty, decor.size=6.5thin}{dol,cv,ccv}			
			\end{fmfgraph*} 		
		}
		\mkern 55mu
		=
			\parbox{25mm}{ 			
			\begin{fmfgraph*}(100,25) 				
				\fmfpen{0.5thin} 				
				\fmfleft{l1,l2,l3,l4,l5}
				\fmfright{r1,r2,r3}				
				\fmftop{t1,t2,c1,a1,cc1,t3,t4}
				\fmfbottom{b1,b2,c2,a2,cc2,b3,b4}
				\fmf{phantom}{t2,dol,b2}
				\fmf{phantom}{dol,r2}
				\fmf{phantom}{t3,dor,b3}
				\fmf{phantom}{dor,r2}
				\fmf{plain}{c1,l4} 				
				\fmf{plain}{l4,dol} 				
				\fmf{plain}{l2,dol} 				
				\fmf{plain}{c2,l2}
				\fmf{plain}{cc1,r2}
				\fmf{plain}{cc2,r2}
				\fmf{plain}{cc1,dor}
				\fmf{plain}{cc2,dor}
				\fmf{wiggly}{c1,a1,cc1}
				\fmf{wiggly}{c2,a2,cc2}
				\fmfv{decor.shape=circle,decor.filled=shaded, decor.size=6.5thin}{a1,a2}					
				\fmfv{decor.shape=circle,decor.filled=empty, decor.size=6.5thin}{cc1,cc2}
				\fmfv{decor.shape=circle,decor.filled=full, decor.size=6.5thin}{c1,c2,dol}			
			\end{fmfgraph*} 		
		}
		\mkern 50mu = \frac{1}{4} \int_{x\neq y\neq z\neq x}\ks^{x}\J_{xy}^{2}\ks^{y}\J_{xz}\kv^{z}\J_{zy}	
	\end{eqnarray} 
\end{fmffile}
In total, the sum of this group of diagrams vanishes. We will demonstrate later that this can be seen as a special case of a diagram bearing a caterpillar subdiagram, which is connected at two points to another subdiagram of a different topology. We will show that the contribution from diagrams of this kind are canceled, akin to the cancelation of contributions from two-particle reducible diagrams for expansions around Gaussian theories.\\

Considering pure ring diagrams, however, we observe that the contributions
\begin{fmffile}{Ring_diagrams} 	
	\begin{eqnarray}		 		
		\fl &-\parbox{25mm}{ 			
			\begin{fmfgraph*}(40,25) 				
				\fmfpen{0.5thin} 				
				\fmfleft{l1,l2,l3,l4,l5}
				\fmfright{r1,r2,r3,r4,r5}				
				\fmftop{t1,t2,cv,t3,t4}
				\fmfbottom{b1,b2,ccv,b3,b4}
				\fmf{phantom}{t2,dol,b2}
				\fmf{phantom}{dol,r3}
				\fmf{phantom}{t3,dor,b3}
				\fmf{phantom}{dor,l3}
				\fmf{plain}{cv,l4} 				
				\fmf{plain}{l4,dol} 				
				\fmf{plain}{l2,dol} 				
				\fmf{plain}{ccv,l2}
				\fmf{plain}{cv,r4}
				\fmf{plain}{r4,dor}
				\fmf{plain}{dor,r2}
				\fmf{plain}{r2,ccv}
				\fmfv{decor.shape=circle,decor.filled=empty, decor.size=6.5thin}{cv,ccv,dol,dor}			
			\end{fmfgraph*} 		
		}
		\mkern -30mu
		+
		\parbox{30mm}{ 			
			\begin{fmfgraph*}(80,25) 				
				\fmfpen{0.5thin} 				
				\fmfleft{l1,l2,l3,l4,l5}
				\fmfright{r1,r2,r3,r4,r5}				
				\fmftop{t1,t2,c1,a1,cv,t3,t4}
				\fmfbottom{b1,b2,c2,a2,ccv,b3,b4}
				\fmf{phantom}{t2,dol,b2}
				\fmf{phantom}{dol,r3}
				\fmf{phantom}{t3,dor,b3}
				\fmf{phantom}{dor,l3}
				\fmf{plain}{c1,l4} 				
				\fmf{plain}{l4,dol} 				
				\fmf{plain}{l2,dol} 				
				\fmf{plain}{c2,l2}
				\fmf{plain}{cv,r4}
				\fmf{plain}{r4,dor}
				\fmf{plain}{dor,r2}
				\fmf{plain}{r2,ccv}				
				\fmf{wiggly}{c1,a1,cv} 				
				\fmf{wiggly}{c2,a2,ccv}
				\fmfv{decor.shape=circle,decor.filled=shaded, decor.size=6.5thin}{a1,a2} 				
				\fmfv{decor.shape=circle,decor.filled=full, decor.size=6.5thin}{dol,c1,c2}
				\fmfv{decor.shape=circle,decor.filled=empty, decor.size=6.5thin}{cv,ccv,dor}			
			\end{fmfgraph*} 		
		}
		-
		\parbox{35mm}{ 			
			\begin{fmfgraph*}(100,25) 				
				\fmfpen{0.5thin} 				
				\fmfleft{l1,l2,l3,l4,l5}
				\fmfright{r1,r2,r3,r4,r5}				
				\fmftop{t1,t2,c1,a1,cv,aa1,cc1,t3,t4}
				\fmfbottom{b1,b2,c2,a2,ccv,aa2,cc2,b3,b4}
				\fmf{phantom}{t2,dol,b2}
				\fmf{phantom}{dol,r3}
				\fmf{phantom}{t3,dor,b3}
				\fmf{phantom}{dor,l3}
				\fmf{plain}{c1,l4} 				
				\fmf{plain}{l4,dol} 				
				\fmf{plain}{l2,dol} 				
				\fmf{plain}{c2,l2}
				\fmf{plain}{cc1,r4}
				\fmf{plain}{r4,dor}
				\fmf{plain}{dor,r2}
				\fmf{plain}{r2,cc2}				
				\fmf{wiggly}{c1,a1,cv,aa1,cc1} 				
				\fmf{wiggly}{c2,a2,ccv,aa2,cc2}
				\fmfv{decor.shape=circle,decor.filled=shaded, decor.size=6.5thin}{a1,a2,aa1,aa2} 				
				\fmfv{decor.shape=circle,decor.filled=full, decor.size=6.5thin}{dol,c1,c2,cc1,cc2,dor}
				\fmfv{decor.shape=circle,decor.filled=empty, decor.size=6.5thin}{cv,ccv}			
			\end{fmfgraph*} 		
		}\\[5mm]
		 \fl = &-\frac{1}{8}\int_{x\neq y \neq u \neq v \neq x}\J_{xu}\kv^{u}\J_{uy}\kv^{y}\J_{yv}\kv^{v}\J_{vx}\kv^{x} \label{eq:Translation_pure_ring_fourth_order}\\
		  \fl &+\frac{1}{2}\int_{x\neq y}\int_{u,v\left(\neq x,y\right)}\J_{xu}\kv^{u}\J_{uy}\kv^{y}\J_{yv}\kv^{v}\J_{vx}\kv^{x}\label{eq:Translation_pure_ring_fourth_order_first_compensating}\\
		  \fl &-\frac{1}{4}\int_{x\neq y}\int_{u,v\left(\neq x,y\right)}\J_{xu}\kv^{u}\J_{uy}\kv^{y}\J_{yv}\kv^{v}\J_{vx}\kv^{x}\label{eq:Translation_pure_ring_fourth_order_second_compensating} 
	\end{eqnarray} 
\end{fmffile}
do not cancel out in general\footnote{If the integration variable is discrete, as for a graph, the sets for which some of them agree do not have vanishing measure and have to be considered separately. We will see in section \ref{Example_spin_systems} that for these special cases, the contribution of the ring diagrams indeed vanishes if exactly one pair of indices is identical.}. This is not a contradiction to the rule concerning diagrams containing caterpillar parts we foreshadowed before because for ring diagrams, the identification of the two nodes, at which the diagram is reducible and therefore decomposes in two parts, is not unique. In section \ref{sec:effectivepotsimpleliquids}, we shall determine the prefactor of a ring diagram of order $n$.\\

The next group of diagrams of fourth order are again watermelon diagrams, and, together with their translation, they read 
\begin{fmffile}{Cherry_diagrams} 	
	\begin{eqnarray}		 		
		 \fl-\quad \parbox{25mm}{ 			
			\begin{fmfgraph*}(50,25) 				
				\fmfpen{0.5thin} 				
				\fmfleft{l1,l2,l3}
				\fmfright{r1,r2,r3}				
				\fmftop{t1,t2,c1,t3,t4}
				\fmfbottom{b1,b2,c2,b3,b4}
				\fmf{phantom}{t2,dol,b2}
				\fmf{phantom}{dol,l2}
				\fmf{phantom}{t3,dor,b3}
				\fmf{phantom}{dor,r2}
				\fmf{plain}{c1,l2} 				
				\fmf{plain}{c2,l2} 				
				\fmf{plain}{c1,dol} 				
				\fmf{plain}{c2,dol}
				\fmf{plain}{c1,r2}
				\fmf{plain}{c2,r2}
				\fmf{plain}{c1,dor}
				\fmf{plain}{c2,dor}
				\fmfv{decor.shape=circle,decor.filled=empty, decor.size=6.5thin}{c1,c2}						
			\end{fmfgraph*} 		
		}
		\mkern -15mu
		&+
		\mkern 5mu
		\parbox{25mm}{ 			
			\begin{fmfgraph*}(75,25) 				
				\fmfpen{0.5thin} 				
				\fmfleft{l1,l2,l3}
				\fmfright{r1,r2,r3}				
				\fmftop{t1,t2,c1,a1,cc1,t3,t4}
				\fmfbottom{b1,b2,c2,a2,cc2,b3,b4}
				\fmf{phantom}{t2,dol,b2}
				\fmf{phantom}{dol,l2}
				\fmf{phantom}{t3,dor,b3}
				\fmf{phantom}{dor,r2}
				\fmf{plain}{c1,l2} 				
				\fmf{plain}{c2,l2} 				
				\fmf{plain}{c1,dol} 				
				\fmf{plain}{c2,dol}
				\fmf{plain}{cc1,r2}
				\fmf{plain}{cc2,r2}
				\fmf{plain}{cc1,dor}
				\fmf{plain}{cc2,dor}				
				\fmf{wiggly}{c1,a1,cc1} 				
				\fmf{wiggly}{c2,a2,cc2}
				\fmfv{decor.shape=circle,decor.filled=shaded, decor.size=6.5thin}{a1,a2} 				
				\fmfv{decor.shape=circle,decor.filled=full, decor.size=6.5thin}{c1,c2}
				\fmfv{decor.shape=circle,decor.filled=empty, decor.size=6.5thin}{cc1,cc2}						
			\end{fmfgraph*} 		
		}
		\mkern 10mu
		&-
		\parbox{25mm}{ 			
			\begin{fmfgraph*}(100,25) 				
				\fmfpen{0.5thin} 				
				\fmfleft{l1,l2,l3}
				\fmfright{r1,r2,r3}				
				\fmftop{t1,t2,c1,a1,cv,aa1,cc1,t3,t4}
				\fmfbottom{b1,b2,c2,a2,ccv,aa2,cc2,b3,b4}
				\fmf{phantom}{t2,dol,b2}
				\fmf{phantom}{dol,l2}
				\fmf{phantom}{t3,dor,b3}
				\fmf{phantom}{dor,r2}
				\fmf{plain}{c1,l2} 				
				\fmf{plain}{c2,l2} 				
				\fmf{plain}{c1,dol} 				
				\fmf{plain}{c2,dol}
				\fmf{plain}{cc1,r2}
				\fmf{plain}{cc2,r2}
				\fmf{plain}{cc1,dor}
				\fmf{plain}{cc2,dor}				
				\fmf{wiggly}{c1,a1,cv,aa1,cc1} 				
				\fmf{wiggly}{c2,a2,ccv,aa2,cc2}
				\fmfv{decor.shape=circle,decor.filled=shaded, decor.size=6.5thin}{a1,a2,aa1,aa2} 				
				\fmfv{decor.shape=circle,decor.filled=full, decor.size=6.5thin}{c1,c2,cc1,cc2}
				\fmfv{decor.shape=circle,decor.filled=empty, decor.size=6.5thin}{cv,ccv}			
			\end{fmfgraph*} 		
		}
		\mkern 58mu\\[5mm]
		\fl =  -\frac{1}{4!} \left(\frac{1}{2}\right)^{4} \cdot 2^{3} \cdot \int_{x\neq y}\J_{xy}^{4}\kk^{x}\kk^{y} 
		    &+ \frac{1}{8}\int_{x\neq y}\frac{\ks^{x}\ks^{x}}{\kv^{x}}\J_{xy}^{4}\frac{\ks^{y}\ks^{y}}{\kv^{y}}
		    &- \frac{1}{16}\int_{x\neq y}\frac{\ks^{x}\ks^{x}}{\kv^{x}}\J_{xy}^{4}\frac{\ks^{y}\ks^{y}}{\kv^{y}}\\
		\fl = 	-\frac{1}{4!} \left(\frac{1}{2}\right)^{4} \cdot 2^{3} \cdot \int_{x\neq y}\J_{xy}^{4}\kk^{x}\kk^{y}
		    &+ \frac{1}{16}\int_{x\neq y}\frac{\ks^{x}\ks^{x}}{\kv^{x}}\J_{xy}^{4}\frac{\ks^{y}\ks^{y}}{\kv^{y}}\label{eq:Cherry_evaluated_general}
	\end{eqnarray} 
\end{fmffile}
Here, the hatched blob associated to the second derivative of $\Gamma_{0}$, unlike the first two groups of diagrams, is attached to cumulants of order higher than two at both sides. Therefore, in general, the first two diagrams are translated in a different way than the last one. However, even if all cumulants of the unperturbed theory are equal, as is the case  for an ideal gas, the contributions from watermelon diagrams do not cancel. For this special case, we will determine the prefactor of a watermelon diagram of order $n$ in section~\ref{sec:effectivepotsimpleliquids}.\\

Finally, there are "glasses" diagrams (as they were christened in \cite{Kuehn18_375004}), which are generated by both line \prettyref{eq:iteration_1pr_remove} and line \prettyref{eq:iteration_2pr_remove} to yield
\begin{fmffile}{Glasses_diagrams}
	\begin{eqnarray}
    	\label{Fourth_order_extraordinary}
			\fl -\mkern 15mu \parbox{25mm}{
				\begin{fmfgraph*}(75,25)
					\fmfpen{0.5thin}
					\fmftop{o1,o2,o3,o4,o5,o6,o7}
					\fmfbottom{u1,u2,u3,u4,u5,u6,u7}
					\fmfright{r1}
					\fmfleft{l1}
					\fmf{phantom}{v3,r1}
					\fmf{phantom}{l1,v1}
					\fmf{phantom}{u1,v1,o3}
					\fmf{plain}{v1,o3}
					\fmf{phantom}{o1,v1,u3}
					\fmf{plain}{v1,u3}
					\fmf{plain}{u3,v2,o5}
					\fmf{plain}{o3,v2,u5}
					\fmf{phantom}{u5,v3,o7}
					\fmf{plain}{u5,v3}
					\fmf{phantom}{o5,v3,u7}
					\fmf{plain}{o5,v3}
					\fmfv{decor.shape=circle,decor.filled=empty, decor.size=6.5thin}{v1,v2,v3}
				\end{fmfgraph*}
				}
				\mkern -9mu
				&+
				\mkern-18mu
				\parbox{25mm}{
				\begin{fmfgraph*}(100,25)
					\fmfpen{0.5thin}
					\fmftop{o1,o2,o3,o4,o5,o6,o7,o8,o9}
					\fmfbottom{u1,u2,u3,u4,u5,u6,u7,u8,u9}
					\fmfright{r1}
					\fmfleft{l1}
					\fmf{phantom}{v3,r1}
					\fmf{phantom}{l1,v1}
					\fmf{phantom}{u1,v1,o3}
					\fmf{plain}{v1,o3}
					\fmf{phantom}{o1,v1,u3}
					\fmf{plain}{v1,u3}
					\fmf{phantom}{u3,v2,o5}
					\fmf{plain}{u3,v2}
					\fmf{phantom}{o3,v2,u5}
					\fmf{plain}{o3,v2}
					\fmf{phantom}{u5,v4,o7}
					\fmf{plain}{v4,o7}			
					\fmf{phantom}{u7,v4,o5}
					\fmf{plain}{u7,v4}
					\fmf{phantom}{u7,v3,o9}
					\fmf{plain}{u7,v3}
					\fmf{phantom}{o7,v3,u9}
					\fmf{plain}{o7,v3}
					\fmf{plain}{v2,g1}
					\fmf{wiggly}{g1,vv1}
					\fmf{wiggly}{vv1,g2}
					\fmf{plain}{g2,v4}
					\fmfv{decor.shape=circle,decor.filled=empty, decor.size=6.5thin}{v1,v2}
					\fmfv{decor.shape=circle,decor.filled=full, decor.size=6.5thin}{v3,v4}
					\fmfv{decor.shape=circle,decor.filled=shaded, decor.size=6.5thin}{vv1}
				\end{fmfgraph*}
				}
				\mkern 27mu
				&-
				\mkern-18mu
				\parbox{25mm}{
				\begin{fmfgraph*}(125,25)
					\fmfpen{0.5thin}
					\fmftop{o1,o2,o3,o4,o5,o6,o7,o8,o9,o10,o11}
					\fmfbottom{u1,u2,u3,u4,u5,u6,u7,u8,u9,u10,u11}
					\fmfright{r1}
					\fmfleft{l1}
					\fmf{phantom}{v3,r1}
					\fmf{phantom}{l1,v1}
					\fmf{phantom}{u1,v1,o3}
					\fmf{plain}{v1,o3}
					\fmf{phantom}{o1,v1,u3}
					\fmf{plain}{v1,u3}
					\fmf{phantom}{u3,v2,o5}
					\fmf{plain}{u3,v2}
					\fmf{phantom}{o3,v2,u5}
					\fmf{plain}{o3,v2}
					\fmf{phantom}{u7,v4,o9}
					\fmf{plain}{v4,o9}			
					\fmf{phantom}{u9,v4,o7}
					\fmf{plain}{u9,v4}
					\fmf{phantom}{u9,v3,o11}
					\fmf{plain}{u9,v3}
					\fmf{phantom}{o9,v3,u11}
					\fmf{plain}{o9,v3}
					\fmf{plain}{v2,g1}
					\fmf{wiggly}{g1,vv1}
					\fmf{wiggly}{vv1,g2}
					\fmf{plain}{g2,v5}
					\fmf{plain}{v5,g3}
					\fmf{wiggly}{g3,vv2}
					\fmf{wiggly}{vv2,g4}
					\fmf{plain}{g4,v4}
					\fmfv{decor.shape=circle,decor.filled=empty, decor.size=6.5thin}{v5}
					\fmfv{decor.shape=circle,decor.filled=full, decor.size=6.5thin}{v1,v2,v3,v4}
					\fmfv{decor.shape=circle,decor.filled=shaded, decor.size=6.5thin}{vv1,vv2}
				\end{fmfgraph*}
				}\\ \nonumber
				\fl = -\frac{3\cdot 2^{4}}{4!2^{4}} \cdot\int_{x\neq y\neq z} \kv^{x} \J_{xy}^{2} \kk^{y} \J_{yz}^{2} \kv^{z}
				&+ \frac{8}{2!2^{4}} \cdot \int_{x\neq y\neq z} \kv^{x} \J_{xy}^{2} \frac{\ks^{y}\ks^{y}}{\kv^{y}} \J_{yz}^{2} \kv^{z}
				&- \frac{4}{2!2^{4}} \cdot \int_{x\neq y\neq z} \kv^{x} \J_{xy}^{2} \frac{\ks^{y}\ks^{y}}{\kv^{y}} \J_{yz}^{2} \kv^{z}\\
				\fl = -\frac{1}{8} \cdot\int_{x\neq y\neq z} \kv^{x} \J_{xy}^{2} \kk^{y} \J_{yz}^{2} \kv^{z}
				  & +\frac{1}{8} \cdot \int_{x\neq y\neq z} \kv^{x} \J_{xy}^{2} \frac{\ks^{y}\ks^{y}}{\kv^{y}} \J_{yz}^{2} \kv^{z} \label{eq:Glasses_diagrams}
	\end{eqnarray}	
\end{fmffile}
For the ideal gas, again, all of these contributions are equal and their sum therefore vanishes. Of course, in general $\frac{\ks^{y}\ks^{y}}{\kv^{y}}\neq \kk^{y}$ and therefore, this is not true for all systems. We now examine how the expansion we have set up unfolds for Ising or Potts spins living on a graph.

\section{Translation of diagrams for Ising and Potts model}\label{Example_spin_systems}
We now work with a spin system where $\psi_x=0,1,\ldots, q-1$ (or $\psi_x=\pm 1$ in the Ising case) and $x$ is a node of a graph. For the diagrams up to third order and the first group of diagrams of fourth order, the evaluation of diagrams is a straightforward application using, of course, the correct expressions for the unperturbed cumulants.

\subsection{Ising model}
For concreteness, we start with the Ising model with the goal of making contact with the approach of \cite{Sessak09}, for which the unperturbed cumulants read
\begin{eqnarray}
	\km^{x}	&:= m_{x}:=\tanh\left(h_{x}\right)\\
	\kv^{x}	&=	\left(1-m_{x}^{2}\right)\\
	\ks^{x}	&=	-2m_{x}\left(1-m_{x}^{2}\right)\\
	\kk^{x}	&=	-2\left(1-3m_{x}^{2}\right)\left(1-m_{x}^{2}\right).
\end{eqnarray}
This yields for the second and third order, evaluating Eq.~\prettyref{eq:Second_order_correction} using Eqs.~\prettyref{eq:Def_coupling}, \prettyref{eq:Def_third_order_ring_diagram} and \prettyref{eq:Def_third_order_Citroen_diagram},
\begin{eqnarray}
	\fl \Gamma^{\mathrm{Ising}}\left[\phi,\Delta\right] =  &\Gamma^{\mathrm{Ising}}_{0}\left[\phi\right] + \frac{1}{4}\sum_{x\neq y}\J_{xy}^{2} \left(1-m_{x}^{2}\right)\left(1-m_{y}^{2}\right)\\
	\fl					&-\frac{1}{6} \sum_{x\neq y\neq z\neq x} \J_{xy} \J_{yz} \J_{zx}\left(1-m_{x}^{2}\right)\left(1-m_{y}^{2}\right)\left(1-m_{z}^{2}\right)\\ 
	\fl				     &-\frac{1}{12} \sum_{x\neq y} \J_{xy}^{3} \left(-2m_{x}\right)\left(1-m_{x}^{2}\right)\left(-2m_{y}\right)\left(1-m_{y}^{2}\right) + \mathcal{O}\left(\Delta^{4}\right),
\end{eqnarray}
which is in agreement with the first two lines of Eq.~(17) in \cite{Sessak09} (they have the opposite sign convention in the definition of the effective potential).\\

To fourth order, specifying Eq.~\prettyref{eq:Translation_pure_ring_fourth_order}, we have to consider three cases depending on the number of identical indices:
\begin{eqnarray}
	\fl &-\frac{1}{8}\int_{x\neq y \neq u \neq v \neq x}\J_{xu}\kv^{u}\J_{uy}\kv^{y}\J_{yv}\kv^{v}\J_{vx}\kv^{x}\\
	\fl = &-\frac{1}{8}\sum_{x,y,u,v\ \mathrm{pairw.} \ \mathrm{uneq.}}\J_{xu}\left(1-m_{u}^{2}\right)\J_{uy}\left(1-m_{y}^{2}\right)\J_{yv}\left(1-m_{v}^{2}\right)\J_{vx}\left(1-m_{x}^{2}\right)\\
	\fl &-\frac{1}{4}\sum_{x,y,u\ \mathrm{pairw.} \ \mathrm{uneq.}}\J_{xu}\left(1-m_{u}^{2}\right)\J_{uy}\left(1-m_{y}^{2}\right)\J_{yu}\left(1-m_{u}^{2}\right)\J_{ux}\left(1-m_{x}^{2}\right)\\
	\fl &-\frac{1}{8}\sum_{x,y\ \mathrm{pairw.} \ \mathrm{uneq.}}\J_{xy}\left(1-m_{y}^{2}\right)\J_{yx}\left(1-m_{x}^{2}\right)\J_{xy}\left(1-m_{y}^{2}\right)\J_{yx}\left(1-m_{x}^{2}\right),
\end{eqnarray}
where the prefactor in the second line is different because there are two possibilities to choose from: $x=u$ or $y=v$. This contribution is canceled by the other ring contributions of fourth order, Eqs.~\prettyref{eq:Translation_pure_ring_fourth_order_first_compensating} and \prettyref{eq:Translation_pure_ring_fourth_order_second_compensating}. All in all, the ring contribution of fourth order is
\begin{eqnarray}
	\fl \Gs^{\mathrm{Ising}}_{4\mathrm{'th}\ \mathrm{order} \ \mathrm{ring}} = &\frac{1}{8}\sum_{x,y,u,v\ \mathrm{pairw.} \ \mathrm{uneq.}}\J_{xu}\left(1-m_{u}^{2}\right)\J_{uy}\left(1-m_{y}^{2}\right)\J_{yv}\left(1-m_{v}^{2}\right)\J_{vx}\left(1-m_{x}^{2}\right)\nonumber\\
	\fl &-\frac{1}{8}\sum_{x\neq y}\J_{xy}\left(1-m_{y}^{2}\right)\J_{yx}\left(1-m_{x}^{2}\right)\J_{xy}\left(1-m_{y}^{2}\right)\J_{yx}\left(1-m_{x}^{2}\right),\label{eq:Ring_opposing_indices_equal}
\end{eqnarray}
where the plus-sign in the first line comes from the overcompensation of the contribution of the first diagram, Eq.~\prettyref{eq:Translation_pure_ring_fourth_order} by those of the other two diagrams, Eqs.~\prettyref{eq:Translation_pure_ring_fourth_order_first_compensating} and \prettyref{eq:Translation_pure_ring_fourth_order_second_compensating}.\\

The watermelon diagrams, Eq.~\prettyref{eq:Cherry_evaluated_general}, can also be evaluated:
\begin{eqnarray}
	\fl -\frac{1}{48} \sum_{x\neq y}\J_{xy}^{4}\left(-2\right)\left(1-3m_{x}^{2}\right)\left(1-m_{x}^{2}\right)\left(-2\right)\left(1-3m_{y}^{2}\right)\left(1-m_{y}^{2}\right)\\
	\fl	    + \frac{1}{16}\sum_{x\neq y} \J_{xy}^{4} 4m_{x}^{2}\left(1-m_{x}^{2}\right)4m_{y}^{2}\left(1-m_{y}^{2}\right)\\
	\fl = -\frac{1}{48} \sum_{x\neq y}\J_{xy}^{4} \left(1-m_{x}^{2}\right) \left(1-m_{y}^{2}\right) \left[4\left(1-3m_{x}^{2}\right)\left(1-3m_{y}^{2}\right) - 48m_{x}^{2}m_{y}^{2}\right]\\
	\fl = -\frac{1}{12} \sum_{x\neq y}\J_{xy}^{4} \left(1-m_{x}^{2}\right) \left(1-m_{y}^{2}\right) \left[1-3m_{x}^{2}-3m_{y}^{2}-3m_{x}^{2}m_{y}^{2}\right].
\end{eqnarray}
The "glasses" diagrams, Eq.~\prettyref{eq:Glasses_diagrams}, lead to the expressions
\begin{eqnarray}
	\fl &-\frac{1}{8} \cdot\sum_{x\neq y\neq z} \left(1-m_{x}^{2}\right) \J_{xy}^{2} \left[\left(-2\right)\left(1-3m_{y}^{2}\right)\left(1-m_{y}^{2}\right)-4m_{y}^{2}\left(1-m_{y}^{2}\right)\right] \J_{yz}^{2} \left(1-m_{x}^{2}\right)\nonumber\\
	\fl = & \frac{1}{4} \cdot\sum_{x\neq y\neq z} \left(1-m_{x}^{2}\right) \J_{xy}^{2} \left(1-m_{y}^{2}\right)^{2} \J_{yz}^{2} \left(1-m_{x}^{2}\right).
\end{eqnarray} 
Finally, we now combine the contribution from the glasses diagrams with $x=z$ with the contribution from the pairwise identification of opposite indices in the ring diagram, Eq.~\prettyref{eq:Ring_opposing_indices_equal} and the watermelon diagram term to obtain
\begin{eqnarray}
	&-\sum_{x\neq y}\J_{xy}^{4}\left(1-m_{x}^{2}\right)\left(1-m_{y}^{2}\right)\left\{ \frac{1}{12}\left[1-3m_{x}^{2}-3m_{y}^{2}-3m_{x}^{2}m_{y}^{2}\right]\right.\\
	&\left.+\frac{1}{8}\left(\left(1-m_{x}^{2}\right)\left(1-m_{y}^{2}\right)\right)^{2}-\frac{1}{4}\left(1-m_{x}^{2}\right)^{2}\left(1-m_{y}^{2}\right)^{2}\right\}\\
	=&\frac{1}{12}\sum_{x<y}\J_{xy}^{4}\left(1-m_{x}^{2}\right)\left(1-m_{y}^{2}\right)\left\{ 1+3m_{x}^{2}+3m_{y}^{2}+9m_{x}^{2}m_{y}^{2}\right\} . 
\end{eqnarray}
Putting everything together, we obtain
\begin{eqnarray}
	\fl \Gs^{\mathrm{Ising}}_{4\mathrm{-th}\  \mathrm{order}} = &\frac{1}{12}\sum_{x<y}\J_{xy}^{4}\left(1-m_{x}^{2}\right)\left(1-m_{y}^{2}\right)\left( 1+3m_{x}^{2}+3m_{y}^{2}+9m_{x}^{2}m_{y}^{2}\right)\\
	\fl &+\frac{1}{4} \sum_{x\neq y\neq z\neq x} \left(1-m_{x}^{2}\right) \J_{xy}^{2} \left(1-m_{y}^{2}\right)^{2} \J_{yz}^{2} \left(1-m_{x}^{2}\right)\\
	\fl &+\frac{1}{8}\sum_{x,y,u,v\ \mathrm{pairw.} \ \mathrm{uneq.}}\J_{xu}\left(1-m_{u}^{2}\right)\J_{uy}\left(1-m_{y}^{2}\right)\J_{yv}\left(1-m_{v}^{2}\right)\J_{vx}\left(1-m_{x}^{2}\right).\nonumber
\end{eqnarray}
This exactly, and elegantly, reproduces the remaining lines of Eq.~(17) in \cite{Sessak09}.

\subsection{Potts model}
Here, the Hamiltonian reads as
\begin{eqnarray}
	H\left[\boldsymbol{\psi}\right] = -\sum_{x}\sum_{\psi^{\prime}=0,\ldots,q-1} h_{x,\psi^{\prime}} \delta_{\psi_{x},\psi^{\prime}}
					     -\frac{1}{2}\sum_{x\neq y}\sum_{\psi^{\prime}=0,\ldots,q-1} J_{xy,\psi^{\prime}} \delta_{\psi_{x},\psi^{\prime}} \delta_{\psi_{y},\psi^{\prime}}.
\end{eqnarray}
The source field for every node $x$ and the coupling for a bond $xy$ is now a $q$-dimensional vector. Therefore, we have to sum not only over the nodes of the graph, but also over the flavors from $0$ to $q-1$ to evaluate a diagram. To do this, we obviously also have to know the expressions for the cumulants in the the uncoupled case. Here, all moments of a spin $\psi_x$ are equal and given by the value of its source field $h_{x}$ through
\begin{eqnarray}
	\fl \km^{x,\psi}=\frac{e^{h_{x,\psi}}}{\sum_{\psi^{\prime}=0,\dots,q-1}e^{h_{x,\psi^{\prime}}}} =: m_{x,\psi}
\end{eqnarray}
and therefore 
\begin{eqnarray}
	\fl \kv^{x,\psi\psi} = m_{x,\psi}\left(1 - m_{x,\psi}\right)\\
	\fl \ks^{x,\psi\psi\psi} = m_{x,\psi} \left(1 - m_{x,\psi}\right) \left(1 - 2m_{x,\psi}\right)\\
	\fl \kk^{x,\psi\psi\psi\psi} = m_{x,\psi} \left(1 - m_{x,\psi}\right) \left( 1 - 6m_{x,\psi} + 6m^{2}_{x,\psi}\right).
\end{eqnarray}
These diagonal elements are sufficient to compute $\Gs$ because the interaction is diagonal in the flavor. We finally obtain
\begin{eqnarray}
	\fl &\Gs^{\mathrm{Potts}}\left[\phi,\Delta\right]\\ 
 \fl =&\Gamma^{\mathrm{Potts}}_{0}\left[\phi\right] + \frac{1}{4}\sum_{x\neq y} \sum_{\psi}\J_{xy,\psi}^{2} m_{x,\psi}\left(1-m_{x,\psi}\right)m_{y,\psi}\left(1-m_{y,\psi}\right)\\
	\fl					&-\frac{1}{6} \sum_{x\neq y\neq z\neq x} \sum_{\psi} \J_{xy,\psi} \J_{yz,\psi} \J_{zx,\psi}m_{x,\psi}\left(1-m_{x,\psi}\right)m_{y,\psi}\left(1-m_{y,\psi}\right)m_{z,\psi}\left(1-m_{z,\psi}\right)\\ 
	\fl				     &-\frac{1}{12} \sum_{x\neq y} \sum_{\psi}\J_{xy,\psi}^{3} m_{x,\psi}\left(1-m_{x,\psi}\right)\left(1-2m_{x,\psi}\right) m_{y,\psi}\left(1-m_{y,\psi}\right)\left(1-2m_{y,\psi}\right)\\
	\fl &+\frac{1}{24} \sum_{x \neq y} \sum_{\psi} \J^{4}_{xy,\psi} m_{x,\psi}\left(1-m_{x,\psi}\right) m_{y,\psi}\left(1-m_{y,\psi}\right)\\
	\fl &\times \left[1-3m_{x,\psi}\left(1-m_{x,\psi}\right)-3m_{y,\psi}\left(1-m_{y,\psi}\right) + 9 m_{x,\psi}\left(1-m_{x,\psi}\right)m_{y,\psi}\left(1-m_{y,\psi}\right)\right] \nonumber\\
	\fl &+\frac{1}{4} \sum_{x\neq y\neq z\neq x} \sum_{\psi}m_{x,\psi}\left(1-m_{x,\psi}\right) \J_{xy,\psi}^{2} \left[m_{y,\psi}\left(1-m_{y,\psi}\right)\right]^{2} \J_{yz,\psi}^{2} m_{x,\psi}\left(1-m_{x,\psi}\right)\\
	\fl &+\frac{1}{8}\sum_{x,y,u,v\ \mathrm{pairw.} \ \mathrm{uneq.}}\sum_{\psi}\J_{xu,\psi}m_{u\psi}\left(1-m_{u\psi}\right)\J_{uy,\psi}m_{y,\psi}\left(1-m_{y,\psi}\right)\\
	\fl &\times \J_{yv,\psi}m_{v,\psi}\left(1-m_{v,\psi}\right)\J_{vx,\psi}m_{x,\psi}\left(1-m_{x,\psi}\right). 
\end{eqnarray}
By equating the derivatives of $\Gs^{\mathrm{Potts}}\left[\phi,\Delta\right]$ (or $\Gs^{\mathrm{Ising}}\left[\phi,\Delta\right]$)  with respect to $\phi$ or to $\Delta$ to the one and two-body couplings, one directly obtains the expression we are after.

\section{Translation into the language of simple liquids}
In the case of simple liquids, the relevant degrees of freedom are the positions $\boldsymbol{x}_{i}$ of the $N$ particles making up the system. We may also choose to work in terms of the empirical
density
\[
	\fl \hat{\rho}\left(\boldsymbol{x}\right):=\sum_{i=1}^{N}\delta\left(\boldsymbol{x}-\boldsymbol{x}_{i}\right),
\]
which depends on the continuous variable $\boldsymbol{x}$. The average of this quantity over the particle positions yields the local density $\rho\left(\boldsymbol{x}\right):=\left\langle \hat{\rho}\left(\boldsymbol{x}\right)\right\rangle $. The densities $\hat{\rho}$ and $\rho$ play the role of $\psi$ and $\phi$ in our earlier presentation. When the particles interact via a pairwise poential $w$ and and are in addition subjected to an external potential $u$, the energy of the system can be expressed as a functional of $\hat{\rho}$ as
\begin{eqnarray}
	\fl E  &= \sum_{i=1}^{N}u\left(\boldsymbol{x}_{i}\right)+\frac{1}{2}\sum_{i\neq j,\,i,j=1}^{N}w\left(\boldsymbol{x}_{i},\boldsymbol{x}_{j}\right)\nonumber\\
	\fl&=\int d\boldsymbol{x}\,\hat{\rho}\left(\boldsymbol{x}\right)u\left(\boldsymbol{x}\right)+\frac{1}{2}\int d\boldsymbol{x}\int d\boldsymbol{y}\,w\left(\boldsymbol{x},\boldsymbol{y}\right)\left[\hat{\rho}\left(\boldsymbol{x}\right)\hat{\rho}\left(\boldsymbol{y}\right)-\hat{\rho}\left(\boldsymbol{x}\right)\delta\left(\boldsymbol{x}-\boldsymbol{y}\right)\right]. 
\end{eqnarray}
The two contributions to $E$ have the form of a one- and a two-point source term for $\hat{\rho}$, respectively, so that the cumulant-generating function of the simple liquid reads, in the grand canonical ensemble with chemical potential $\mu$,
\begin{eqnarray}
	\fl W\left[\nu,w\right]= & \ln\left\{ \sum_{N=0}^{\infty}\frac{1}{N!}\int d\boldsymbol{x}_{1}\dots\int d\boldsymbol{x}_{N}\exp\left(\int d\boldsymbol{x}\,\hat{\rho}\left(\boldsymbol{x}\right)\nu\left(\boldsymbol{x}\right)\right.\right.\\
 	\fl & \left.\left.-\beta\frac{1}{2}\int d\boldsymbol{x}\int d\boldsymbol{y}\,w\left(\boldsymbol{x},\boldsymbol{y}\right)\left[\hat{\rho}\left(\boldsymbol{x}\right)\hat{\rho}\left(\boldsymbol{y}\right)-\hat{\rho}\left(\boldsymbol{x}\right)\delta\left(\boldsymbol{x}-\boldsymbol{y}\right)\right]\right)\right\} ,
\end{eqnarray}
where the one-body source $\nu\left(\boldsymbol{x}\right)=-\beta\left(u\left(\boldsymbol{x}\right)-\mu\right)$ now incorporates the chemical potential. In the absence of an interaction potential, denoting $W_0[\nu]=W[\nu,0]$, we see that the statistics of the empirical density is fully known. The cumulants of $\hat{\rho}$, for instance, are given by:
\begin{eqnarray}
	\fl \langle\left(\hat{\rho}\left(\boldsymbol{x}\right)\right)^{n}\rangle_c  =\frac{\delta^{n}}{\delta\nu\left(\boldsymbol{x}_{1}\right)\dots\delta\nu\left(\boldsymbol{x}_{n}\right)}W_{0}\left[\nu\right]\nonumber  =\exp\left(\nu\left(\boldsymbol{x}_{1}\right)\right)\prod_{i=1}^{n-1}\delta\left(\boldsymbol{x}_{i+1}-\boldsymbol{x}_{i}\right).\label{eq:cumulants_ideal_gas}
\end{eqnarray}
In the presence of interactions, $w\neq 0$, the functional $W[\nu,w]$ can be obtained from $W_0[\nu]$ as
\begin{eqnarray}
	\fl e^{W\left[\nu,w\right]}= & \exp\left(-\frac{\beta}{2}\int d\boldsymbol{x}\int d\boldsymbol{y}\,w\left(\boldsymbol{x},\boldsymbol{y}\right)\left[\frac{\delta}{\delta\nu\left(\boldsymbol{x}\right)}\frac{\delta}{\delta\nu\left(\boldsymbol{y}\right)}-\delta\left(\boldsymbol{x}-\boldsymbol{y}\right)\frac{\delta}{\delta\nu\left(\boldsymbol{x}\right)}\right]\right)e^{W_0[\nu]}.\label{eq:Full_cum_gen_fct}
\end{eqnarray}
This expression is similar in spirit to Eq.~\prettyref{eq:Gamma2_integral_expression} (with the correspondence $\psi \rightarrow \hat{\rho}$, $\phi \rightarrow \rho$ and $i \rightarrow \boldsymbol{x}$), up to the fact that here we are working directly with the cumulant-generating function instead of with the second Legendre transform. Note that the additional $\delta\left(\boldsymbol{x}-\boldsymbol{y}\right)\frac{\delta}{\delta\nu\left(\boldsymbol{x}\right)}$ in the exponential in Eq.~\prettyref{eq:Full_cum_gen_fct} expresses that we exclude self-interactions from the energy $E$.

This connection between the static theory of simple liquids~\cite{hansen2013theory,morita1960new,morita1961new} and field theory has seldom been exploited in the past~\cite{caillol2006statistical,caillol2006non,vasiliev2019functional}. For completeness, in appendix~\ref{sub:Two_Mayer_thms}, we have given the proofs to the two theorems due to Goeppert-Mayer and Mayer~\cite{Mayer77} using the formalism of this chapter. The first of these two theorems, which is valid generically for systems described by Eq.~\prettyref{eq:Def_cumulant_generating_fct}, states that the cumulant-generating function is represented by all connected diagrams. The second Mayer theorem relies on the Poisson statistics of the ideal gas (with all cumulants Eq.~\prettyref{eq:cumulants_ideal_gas} being equal), and it is therefore only valid for simple liquids. It states that the first Legendre transform is represented by all one-particle irreducible diagrams, that is, those that cannot be cut into two pieces both containing interactions by removing a node representing a single cumulant.

\section{Effective potential for simple liquids}\label{sec:effectivepotsimpleliquids}
Combining the results of the previous sections, we have everything at hand to derive a diagrammatic expansion of the second Legendre transform in a simple liquid. This is easier to obtain than its analog in systems with a finite number of states for two reasons. First of all, all cumulants are equal in the unperturbed theory and second, the integration variables are continuous and uniform, and this waives part of the technicalities in which distinguishing equal lattice sites (as in the Ising model) was necessary.\\

When discussing the diagrammatics in the general case, we have identified two classes of diagrams that play an important role in simple liquids. These are the watermelon and ring diagrams. We will show that these, together with two-particle irreducible diagrams, are the only ones contributing to the expansion of the second Legendre transform (our proof is completely different from that of \cite{morita1960new,morita1961new} but perhaps closer in spirit to that of \cite{vasiliev2019functional}). For theories with two-body interactions, two-particle irreducible diagrams are defined by their property that, by removing two nodes, they cannot be separated either into more than two pieces or into two pieces each of which containing more than one interaction. Hence the first nontrivial diagrams to consider are the fourth-order ones.\\

To fourth order, we have noted that the ring and the watermelon diagram, albeit two-particle reducible, do not cancel. This is also true for their higher-order equivalents, the symmetry factors of which, $S_{\mathrm{ring},n}$ and $S_{\mathrm{watermelon},n}$, are given by
\begin{eqnarray}
	S_{\mathrm{ring},n} &= \frac{\left(-1\right)^{n}}{2n} \label{eq:Statement_symmetry_factor_ring}\\ 
	S_{\mathrm{watermelon},n} &= \frac{\left(-1\right)^{n}}{2n\left(n-1\right)}.\label{eq:Statement_symmetry_factor_cherry}
\end{eqnarray}
We now prove both of these statements by induction, assuming that for $n$ arbitrary, but fixed, they hold for all $k<n$. As we have seen in the fourth order, there are several ways to construct a ring diagram of $n$-th order. Concretely, we can cut the ring into several pieces at any node, the only restriction being that the largest part is of order $n-2$ (because, according to the general rules, subdiagrams with $n-1$ interactions must not be used to construct diagrams of order $n$). All of these contributions yield the same expressions, only with different numerical prefactors and signs. To keep track of those, it is convenient to define an index set that we call the reduced partition $P_{r}\left(n\right)$ of $n$
\begin{eqnarray}
	\fl P_{r}\left(n\right)=\left\{ \boldsymbol{k}\in\mathbb{N}_{0}^{n-2}|n=\sum_{i=1}^{n-2}k_{i}i, \, \sum_{i} k_{i} > 0\right\}\label{eq:Def_reduced_partition_fct}.
\end{eqnarray}
Let $\boldsymbol{k}\in P_{r}\left(n\right)$ characterize a ring diagram of order $n$ decomposed into $k_{1}$ pieces of length $1$, $k_{2}$ pieces of order $2$ and so on. There are $\left(\sum_{i} k_{i}\right)!$ possibilities to concatenate them into a "caterpillar". Sticking the ends of it together to form a ring, we reduce this symmetry factor by an amount $\frac{1}{2\sum_{i} k_{i}}$ because we can start to concatenate at every node and we can go either clockwise or counter-clockwise. Then, taking the derivative of a ring diagram of order $i+1$, we obtain a caterpillar of order $i$ bearing the prefactor $\left(-1\right)^{i+1}$ because there are $i+1$ pairwise correlations to differentiate, which are symmetric, hence the $\frac{1}{2}$ is compensated as well. Combining this with the prefactor $\frac{1}{k_i!}$ for every group of caterpillars, summing up the symmetry factors of all contributions and taking the global minus sign into account, we eventually obtain
\begin{eqnarray}
	 \fl S_{\mathrm{ring},n} = -\sum_{\boldsymbol{k}\in P_{r}\left(n\right)}\frac{\left(\sum_{i=1}^{n-2}k_{i}-1\right)!}{2}\prod_{i=1}^{n-2}\frac{\left(-1\right)^{\left(i+1\right)k_{i}}}{k_{i}!}.
\end{eqnarray}
By a lengthy, yet elementary calculation, we show in appendix \ref{Number_theory_isolated_ring} that this expression is actually identical to Eq.~\prettyref{eq:Statement_symmetry_factor_ring}. We proceed similary for the watermelon diagrams, in particular, we use the same partitions of $n$. Differentiating the contribution of order $i+1, i\leq n-2$ with respect to $\Delta$, we obtain, according to the induction assumption, a factor $\frac{\left(-1\right)^{i+1}}{i}$. Besides the flipping of vertices, there are no further symmetries to consider so that we obtain
\begin{eqnarray}
	\fl S_{\mathrm{watermelon},n} = -\sum_{\boldsymbol{k}\in P_{r}\left(n\right)}\frac{1}{2}\prod_{i=1}^{n-2}\frac{\left(-1\right)^{\left(i+1\right)k_{i}}}{k_{i}!i^{k_{i}}}.
\end{eqnarray}
and by a calculation similar to that of the ring diagrams (details in appendix~\ref{Number_theory_isolated_cherry}), we arrive at Eq.~\prettyref{eq:Statement_symmetry_factor_cherry}.\\ 

\begin{figure}
	\center
	\includegraphics[width=.4\textwidth]{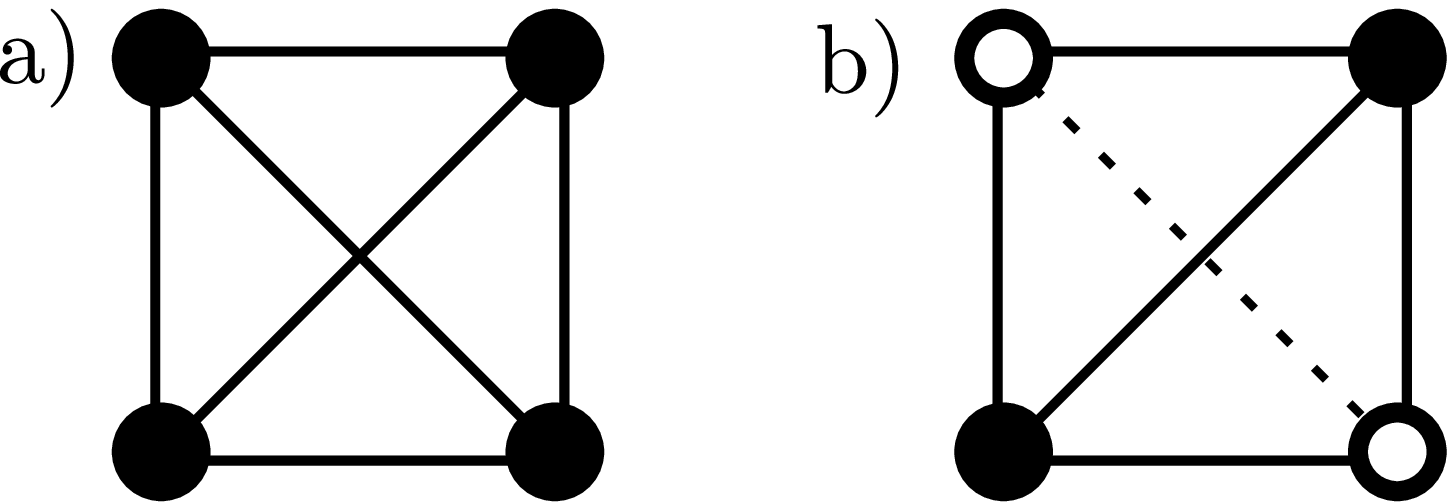}
	\caption{a) The two-particle irreducible diagram of lowest order, the "tetraeder", b) with one interaction removed.}
	\label{fig:tetraeder_full_and_with_one_line_removed}	
\end{figure}

What are the other diagrams contributing to $\Gs$? First, there are diagrams that are two-particle irreducible, which can only be built sticking \J's together. These are rather rare, the first of them is of sixth order, in which four nodes are connected all-to-all by interactions (termed "tetraeder" in \cite{vasiliev2019functional}, see \prettyref{fig:tetraeder_full_and_with_one_line_removed}).\\
What about two-particle reducible diagrams other than those of ring or watermelon type? We can discriminate the following cases here:
\begin{enumerate}
	\item The $\Delta$-derivative of a two-particle irreducible subdiagram is stuck to the remainder at two nodes;
	\item One caterpillar diagram is attached to the remainder at two nodes\label{Caterpillar_case_2pr};
	\item One watermelon diagram is attached to the remainder at two nodes\label{Cherry_case_2pr}.
\end{enumerate}
We will prove by induction that all three contributions vanish. We can therefore assume in the following that all diagrams of lower order than the one we are constructing are either two-particle irreducible or of the ring or the watermelon type.\\
Considering the first case, we observe that by differentiating a two-particle irreducible diagram with respect to $\Delta$, in general we obtain a two-particle reducible diagram. However, it is impossible that the differentiated diagram decomposes into two pieces by removing the nodes the removed interaction was attached to. Indeed, diagrammatically, this would look like
\begin{fmffile}{Explanation_undangerously_2pr}
	\begin{eqnarray}
	\parbox{30mm}{
		\begin{fmfgraph*}(70,30)
			\fmfpen{0.5thin}
			\fmfleft{l}
			\fmfright{r}
			\fmfbottom{b}
			\fmftop{t}
			\fmf{plain}{l,t,r,b,l}
			\fmf{dashes}{t,b}
			\fmfv{decor.shape=circle,decor.filled=empty, decor.size=6.5thin}{t,b}
			\fmfv{decor.shape=circle,decor.filled=hatched, decor.size=10thin}{l,r}
			\end{fmfgraph*}
	},
	\end{eqnarray}
\end{fmffile}
where the hatched blobs represent the remainder of the diagram and the dotted line the removed interaction. We see that if the differentiated diagram were two-particle reducible at these nodes, the original diagram would be two-particle reducible as well. This is a contradiction to our assumption. Let us consider the lowest-order two-particle irreducible diagram, shown in \prettyref{fig:tetraeder_full_and_with_one_line_removed}, as an example to illustrate this statement: Removing any interaction from the tetraeder, we obtain a two-particle reducible diagram. We might therefore fear that, besides at the two points we join the other subdiagrams, we obtain two other points at which the whole diagram can be decomponsed into two nontrivial pieces. Indeed, as shown in panel b) of \prettyref{fig:tetraeder_full_and_with_one_line_removed}, the tetraeder with one interaction removed decomposes even into four parts by removing the other diagonal interaction (and the cumulants attached to it). However, it is exactly the interaction  that is {\em not} attached to the interaction removed. During the iterative generation of diagrams, we are thus not allowed to connect any subdiagram at these places (which would add a two-particle reducible diagram in the higher order). We see with this example that the only points, at which a diagram generated during the iteration can be two-particle reducible, are exactly the joining points of the subdiagrams. Consequently, subdiagrams built from removing an interaction from a two-particle irreducible diagram can enter a diagram in only two ways: Either by using the derivative of the corresponding two-particle irreducible diagram as interaction or by rebuilding a subdiagram of the same shape with interactions $\J$ of order $1$. That both of these two contributions also contribute with the same symmetry factor can be shown analogously to the proof for two-particle irreducibility of the diagrams for $\Gs$ expanded around a Gaussian theory: Building the subdiagram with $\J$'s, the symmetry factor of the entire diagram can be determined by first choosing $\left(\begin{array}{c}n+n^{\prime}\\n \end{array}\right)$ interactions to be part of the subdiagram we assume to be of order $n$ (the remainder being of order $n^{\prime}$). Together with the prefactor $\frac{1}{\left(n+n^{\prime}\right)!}$, this yields $\frac{1}{n!n^{\prime}!}$, corresponding to the orders of the subdiagrams. Consequently, both ways to construct the diagram yield the same value. Because, however, the subdiagram comes with a minus sign if it is part of a $\Delta$-derivative of $\Gs$, the sum of these contributions cancel. As a final remark for this case, let us note that the factor $2$ coming about by differentiating the two-particle irreducible subdiagram by $\Delta$ is compensated by the factor $\frac{1}{2}$ in front of any two-point term in the differential operator $A$ as defined in ~Eq. \prettyref{eq:Def_A_diff_op}.
\\

\begin{figure}
	\center
	\includegraphics[width=.5\textwidth]{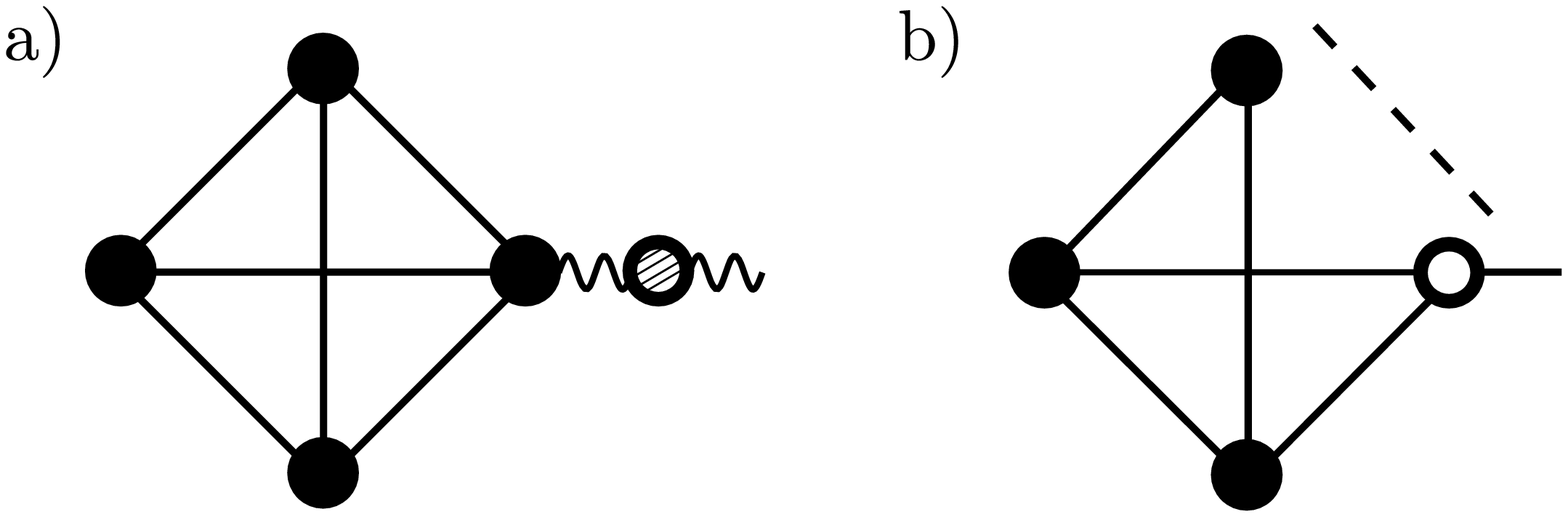}
	\caption{Example of two contributions that together cancel one-particle reducible diagrams.}
	\label{fig:One-particle-reducible_cancelling_blobs}
\end{figure}

Case (\ref{Caterpillar_case_2pr}) is different because, similarly to the construction of ring diagrams, there is more than one possibility to create the caterpillar diagram. Assume that the caterpillar contains $n>1$ interactions and the remainder $n^{\prime}$. As for the ring diagrams, the caterpillars can be built from pieces representing partitions of $n$, with the only difference that here, all partitions
\begin{eqnarray}
	\fl P\left(n\right)=\left\{ \boldsymbol{k}\in\mathbb{N}_{0}^{n}|n=\sum_{i=1}^{n}k_{i}i,\, \sum_{i}k_{i}>0\right\}.
\end{eqnarray}
are allowed. Furthermore, now both the starting point and the direction of the concatenation matter so that the sum of the symmetry factors of all of these contributions is
\begin{eqnarray}
	\fl \sum_{\boldsymbol{k}\in P\left(n\right)}\left(\sum_{i=1}^{n}k_{i}\right)!\prod_{i=1}^{n}\frac{\left(-1\right)^{\left(i+1\right)k_{i}}}{k_{i}!}.
\end{eqnarray}
In appendix~\ref{Number_theory_caterpillars}, we show, by slightly adapting the calculation of appendix~\ref{Number_theory_isolated_ring}, that this expression vanishes unless $n=1$. Because in caterpillars, all cumulants are of order $2$, the statement that diagrams with integrated caterpillars yield vanishing contributions, holds true for all theories. It can also be used to prove that the first group of diagrams of fourth order we discussed in subsection \ref{subsec:terramystica} cancel.\\

For case (\ref{Cherry_case_2pr}), we argue in a similar fashion; also there, the only adaptation is switching from the reduced to the full partition, this time for constructing all watermelon-subdiagrams instead of ring diagrams. The sum of the symmetry factors of all watermelon subdiagrams (which amount to multiple connections between two nodes) is then
\begin{equation}
	\fl \sum_{\boldsymbol{k}\in P\left(n\right)}\prod_{i=1}^{n}\frac{\left(-1\right)^{\left(i+1\right)k_{i}}}{k_{i}!i^{k_{i}}}.
\end{equation}
In appendix~\ref{Number_theory_multiple_connections}, we demonstrate that this expression also vanishes unless $n=1$. We have therefore shown that the diagrams of cases i)-iii) indeed cancel out.\\

Finally, we have to prove that the same applies to all one-particle reducible diagrams. In contrast to, {\it e.g.}, a Gaussian $\phi^{4}$-theory, the interaction depends on $\hat{\rho}$: $\J_{xy} \sim \frac{1}{\hat{\rho}_{x}\hat{\rho}_{y}}$. Thus, each circle attached to $n$ lines represents a quantity $\propto \frac{1}{\hat{\rho}^{n-1}}$ and its derivative is proportional to $-\left(n-1\right) \frac{1}{\hat{\rho}^{n}}$. It might seem as if an additional prefactor is introduced and therefore sticking it together with another subdiagram does not lead to the same symmetry factor as the same diagram created by interactions of order $1$. However, there is yet another possibility to obtain the same subdiagram than by differentiating with respect to $\hat{\rho}$, namely removing one of the attached lines by a $\Delta$-derivative and replacing it with a first-order interaction. Note that here, we are discussing a diagram that is composed of a subdiagram of order $n$ attached to another diagram of order $n'$. Constructing the isolated order $n$ diagram we would only be allowed to use subdiagrams of order $n-2$. However, subdiagrams of order $n-1$ contribute once this diagram is integrated into a larger one. These possibilities are depicted in \prettyref{fig:One-particle-reducible_cancelling_blobs}. In (b), because we have $n$ lines to potentially replace, the diagram translates with a prefactor $n$. Adding both contributions, we thus obtain the prefactor $1$. Because the two-particle irreducible diagram that we differentiate contributes with a minus sign, this amounts the cancellation of one-particle reducible diagrams - that the remaining symmetry factors are the same for all variants to build a diagram of a certain topology can again be argued analogously to the Gaussian case.\\

This completes the characterization of the diagrams contributing to $\Gs$ for the case in which the unperturbed theory is the ideal gas: They are either of the ring or the watermelon type, or they are two-particle irreducible.

\section{Outlook}
A short summary of our results is as follows: we have designed some diagrammatics to deal with perturbations around a noninteracting non-Gaussian theory, at fixed field average and correlations. This is thus a perturbative approach in powers of interactions to the inverse problem which consists in finding the couplings able to achieve these two sets of constraints on the field. The diagrammatic approach we have set up is rather powerful, as we have illustrated on two canonical systems: the Ising model (previously approached in \cite{Sessak09}) and simple liquids (also explored by other methods in \cite{morita1960new} in a more distant past). It is also remarkably aesthetic, in our opinion, that such distant problems as the Ising model or the ideal gas, or standard field theories,  can actually be tackled with the exact same formalism. Of course, this proof of concept calls for further, yet unexplored, extensions and applications.\\

In terms of extensions, we have worked with a noninteracting (non-Gaussian) theory perturbed by two-point interactions. Certainly, some of these restrictions could be waived. For instance, one could work with a correlated non-Gaussian theory, as long as the correlations of the latter are available. Of course, higher-order interactions could also be considered.\\

In terms of physical applications, we see several interesting problems. In the theory of simple liquids the statics of the ideal gas is not described by a Gaussian theory. Dynamics-wise, the same holds: the description of the dynamics of an assembly of, say, independent Brownian particles, involves a dynamical action (Martin-Siggia-Rose-Janssen-De Dominicis) which is non-Gaussian in the density field~\cite{velenich2008brownian}. To be specific, it is quadratic in the Doi-Peliti fields, but it is cubic in the density field and its response field partner. This renders approximations based on perturbing around the ideal gas particularly cumbersome. This has for instance been illustrated in a series of papers striving to study the dynamics of the dense phase by field-theoretic methods~\cite{miyazaki2005mode,andreanov2006dynamical,Jacquin11_210602,kim2014equilibrium} (similar issues arise in systems with disorder-mediated interactions~\cite{Kim2020dynamics}). With the tools presented in this work, this opens up the possibility of a perturbation expansion where the non-Gaussian theory already satisfies (exactly) a fluctuation-dissipation theorem. For instance, one could perform a perturbation of the generating functional in powers of the cubic vertex encoding interactions between particles. More recently, the techniques of \cite{Sessak09} have been applied in \cite{Mahuas22small} to neural networks in the sensory system. These authors used the Ising model to compute entropies of the neurons' activities. However it is known that working with two-state degrees of freedom is a limitation (ideally, a larger number of internal states should be used). Finding out the robustness of entropic calculations with respect to the reference model is of course a very interesting question which our formalism could help sort out.\\

Beyond glassy dynamics, where the approach would be used directly to establish an evolution equation for correlations, our tools can also be exploited for actual inference questions. We have in mind, for instance, projecting the data retrieved from the simulation  of a Vicsek model~\cite{vicsek1995novel} (or the observation of some bird flock) onto a Heisenberg model~\cite{bialek2012statistical}. We are perfectly aware that there are other methods to achieve that goal, and that these other methods are numerically more powerful. However, our approach should be able to shed some analytical light on problems that deserve to be attacked by several methods.

\section{Acknowledgements}
We thank Michael Dick and Ulisse Ferrari for valuable feedback on the manuscript.
\section{Appendix}

\subsection{Generalized treatment of Gaussian theories}\label{General_non_Gauss}
In the approach employed in \prettyref{sec:Gaussian_theory_main} we do not treat the one- and two-point functions on equal footing. One could also add a two-point source term to the unperturbed theory, which we choose, together with $j$ so as to make $W_{0}$ yield the right cumulants. Then, one could use $\Gs\left[\phi,\Delta\right] = - g\left[j[\phi,\Delta],K\left[\phi,\Delta\right]\right]$ to obtain the derivatives of $\Gs$, which yields
\begin{eqnarray}
	\fl	\frac{\del\Gs^{\ii}}{\del\phi_{x}} & = \int_{u} \frac{\del j_{u}}{\del\phi_{x}}\frac{\del}{\del j_{u}}g\left[j,K\right] = \left(\frac{\del^{2}\Gs_{0}}{\del\phi^{2}}-\frac{\del\Gs_{0}}{\del\Delta}\right)\frac{\del}{\del j}g\left[j,K\right]\nonumber \\
	\fl \frac{\del\Gs^{\ii}}{\del\Delta_{xy}} & = \frac{1}{2}\int_{u\neq v}\frac{\del^{2}\Gs_{0}}{\del\Delta_{xy}\del\Delta_{uv}}\left(\frac{\del}{\del K_{uv}}-\phi_{u}\frac{\del}{\del j_{v}}-\phi_{v}\frac{\del}{\del j_{u}}\right)g\left[j,K\right],\label{eq:Deriv_Delta_Gamma_Gaussian}
\end{eqnarray}
In the respective last steps, we have used that $\frac{\del^{2}\Gs}{\del \Delta\del\phi}=0$ if the unperturbed theory is Gaussian. In this case, furthermore
\begin{eqnarray}
	\fl \int_{u} \left(\frac{\del^{2}\Gs}{\del\phi_{x}\phi_{u}} - \frac{\del\Gs}{\del\Delta_{xu}}\right) \frac{\del^{2}W}{\del j_{u}\del j_{y}} &= \delta_{xy} \label{eq:Inv_relation_single_prop}\\
	\fl \frac{1}{2}\int_{u\neq v} \frac{\del^{2}\Gs}{\del\Delta_{xu}\del \Delta_{yv}} \frac{\del^{2}W}{\del j_{u} \del j_{x^{\prime}}} \frac{\del^{2}W}{\del j_{v} \del j_{y^{\prime}} } &= \delta_{xx^{\prime}} \delta_{yy^{\prime}}.\label{eq:Inv_relation_double_prop}
\end{eqnarray}
In order to evaluate Eq.~\prettyref{eq:Deriv_Delta_Gamma_Gaussian}, we furthermore have to rewrite the $K$-derivatives in terms of $j$-derivatives. This is necessary because the contributions stemming from the bare interaction are expressed in terms of those. The unperturbed theory being Gaussian, it suffices to compute the $K$-derivative of the first and second derivative of $W_{0}$:
\begin{eqnarray}
	\fl \frac{\del}{\del K_{xy}} \frac{\del W_{0}}{\del j_{u}} &= \frac{\del^{2} W_{0}}{\del j_{x}\del j_{u}} \frac{\del W_{0}}{\del j_{y}} + \frac{\del^{2} W_{0}}{\del j_{y}\del j_{u}} \frac{\del W_{0}}{\del j_{x}} \label{eq:Deriv_K_first_cumulant}\\ 
	\fl \frac{\del}{\del K_{xy}} \frac{\del^{2} W_{0}}{\del j_{u} \del j_{v}} &= \frac{\del^{2} W_{0}}{\del j_{u} \del j_{x}} \frac{\del^{2} W_{0}}{\del j_{v} \del j_{y}} + \frac{\del^{2} W_{0}}{\del j_{u} \del j_{y}} \frac{\del^{2} W_{0}}{\del j_{v} \del j_{x}}
\end{eqnarray}
Comparing Eq.~\prettyref{eq:Deriv_K_first_cumulant} with Eq.~\prettyref{eq:Deriv_Delta_Gamma_Gaussian}, we observe that the terms with $j$-derivatives in Eq.~\prettyref{eq:Deriv_Delta_Gamma_Gaussian} are canceled by the terms coming about by letting the $K$-derivative act on a first-order $j$-derivative (after inserting the right value of $j$). Therefore, Eq.~\prettyref{eq:Deriv_Delta_Gamma_Gaussian} can be represented by a diagram with two external legs. Now, using Eqs.~\prettyref{eq:Inv_relation_single_prop} and \prettyref{eq:Inv_relation_double_prop}, we have reproduced the result of the main text, namely that a $\phi$-derivative corresponds to removing a dangling leg and a $\Delta$-derivative to the removal of a propagator line. 
Consequently, also the result that only two-particle irreducible diagrams contribute to the expansion of $\Gs^{\ii}$ is reproduced. In principle, one could generalize our result to general non-Gaussian theories in this way. However, this is harder to to do than it is for the first Legendre transform because in this case $\frac{\del^{2}\Gamma_{0}}{\del \phi \del \Delta}\neq 0$ in general and therefore, Eqs.~\prettyref{eq:Inv_relation_single_prop} and \prettyref{eq:Inv_relation_double_prop} do not hold anymore.

\subsection{Proving Mayers' theorems}\label{sub:Two_Mayer_thms}
\subsubsection{First Mayer theorem}\label{sub:First_Mayer_thm}
Loosely following the proof presented in \cite[appendix 3]{Kuehn18_375004} and adapting it to the language of simple liquids,
we will demonstrate that $W$ is represented by all connected diagrams,
which is known as the first Mayer theorem. Splitting the cumulant-generating
function into a solvable and a perturbing part as $W=W_{0}+W_{V}$,
we rewrite \prettyref{eq:Full_cum_gen_fct} as
\begin{eqnarray}
	\fl &e^{W_{V}\left[\nu,w\right]}  \\ 
	\fl = &e^{-W_{0}\left[\nu\right]}\lim_{L\rightarrow \infty}\left(1-\frac{1}{L}\frac{\beta}{2}\int_{\boldsymbol{x}}\int_{\boldsymbol{y}}\,w\left(\boldsymbol{x},\boldsymbol{y}\right)\left[\frac{\delta}{\delta\nu\left(\boldsymbol{x}\right)}\frac{\delta}{\delta\nu\left(\boldsymbol{y}\right)}-\delta\left(\boldsymbol{x}-\boldsymbol{y}\right)\frac{\delta}{\delta\nu\left(\boldsymbol{x}\right)}\right]\right)^{L}\int d\boldsymbol{x}\,e^{\nu\left(\boldsymbol{x}\right)}.\nonumber
\end{eqnarray}
Assume now that we apply the perturbing part only $l\leq L$ times
and, accordingly, define
\[
\fl e^{W_{l,L}\left[\nu,w\right]}:=\left(1-\frac{\beta}{L}A\left[\frac{\delta}{\delta\nu}\right]\right)^{l}e^{W_{0}\left[\nu\right]},
\]
where we have abbreviated 
\[
\fl A\left[\frac{\delta}{\delta\nu}\right]:=\frac{1}{2}\int d\boldsymbol{x}\int d\boldsymbol{y}\,w\left(\boldsymbol{x},\boldsymbol{y}\right)\left[\frac{\delta}{\delta\nu\left(\boldsymbol{x}\right)}\frac{\delta}{\delta\nu\left(\boldsymbol{y}\right)}-\delta\left(\boldsymbol{x}-\boldsymbol{y}\right)\frac{\delta}{\delta\nu\left(\boldsymbol{x}\right)}\right]
\]
Then we clearly have that
\[
\fl e^{W_{l+1,L}\left[\nu,w\right]}=\left(1-\frac{\beta}{L}A\left[\frac{\delta}{\delta\nu}\right]\right)e^{W_{l,L}\left[\nu,w\right]}
\]
and therefore
\begin{eqnarray}
	\fl W_{l+1,L}\left[\nu,w\right] & =\ln\left(\left(1-\frac{\beta}{L}A\left[\frac{\delta}{\delta\nu}\right]\right)e^{W_{l,L}\left[\nu,w\right]}\right)\\
 & =-e^{-W_{l,L}\left[\nu,w\right]}\frac{\beta}{L}A\left[\frac{\delta}{\delta\nu}\right]e^{W_{l,L}\left[\nu,w\right]}+\mathcal{O}\left(\frac{1}{L^{2}}\right).
\end{eqnarray}
This tells us that we obtain the contributions of order $l+1$ to
the cumulant-generating function by binding together lower-order ones
by an interaction. Diagrammatically this means that we connect an
interaction vertex to lower-order contributions, including $W_{0}$.
As in the case of the effective potential discussed in the main text, at every iteration step, at most one interaction survives the $L\to+\infty$ limit. 

\subsubsection{Resummations for simple liquids}
We can exploit the perturbation expansion of the simple liquid to perform partial resummations. Consider the sum of all watermelon diagrams:

\begin{fmffile}{Resummation_chevron}	
	\begin{eqnarray}
		\parbox{25mm}{
			\begin{fmfgraph*}(25,25)
				\fmfpen{0.5thin}
				\fmftop{o1,o2,o3}
				\fmfbottom{u1,u2,u3}
				\fmf{plain}{u1,o2}
				\fmf{plain}{u3,o2}
				\fmfv{decor.shape=circle,decor.filled=empty, decor.size=6.5thin}{u1,u3}
			\end{fmfgraph*}
		} 
		\mkern-60mu + \mkern-20mu \parbox{25mm}{
			\begin{fmfgraph*}(75,25)
				\fmfpen{0.5thin}
				\fmftop{o1,o2,o3,o4,o5}
				\fmfbottom{u1,u2,u3,u4,u5}
				\fmf{phantom}{u1,v1,o3}
				\fmf{plain}{v1,o3}
				\fmf{phantom}{o1,v1,u3}
				\fmf{plain}{v1,u3}
				\fmf{phantom}{u3,v2,o5}
				\fmf{plain}{u3,v2}
				\fmf{phantom}{o3,v2,u5}
				\fmf{plain}{o3,v2}
				\fmfv{decor.shape=circle,decor.filled=empty, decor.size=6.5thin}{v1,v2}
			\end{fmfgraph*}
			}
			+
			\mkern20mu \parbox{25mm}{
			\begin{fmfgraph*}(25,25)
				\fmfpen{0.5thin}
				\fmftop{o1,o2,o3}
				\fmfbottom{u1,u2,u3}
				\fmf{phantom}{u2,du,do,o2}
				\fmf{phantom,tension=2}{du,do,o2}
				\fmf{plain}{u1,o2}
				\fmf{plain}{u3,o2}
				\fmf{plain}{u1,do}
				\fmf{plain}{u3,do}
				\fmf{plain}{u1,du}
				\fmf{plain}{u3,du}
				\fmfv{decor.shape=circle,decor.filled=empty, decor.size=6.5thin}{u1,u3}
			\end{fmfgraph*}
		}
		\mkern-50mu
		+ \mkern-30mu 
		\parbox{25mm}{
			\begin{fmfgraph*}(75,25)
				\fmfpen{0.5thin}
				\fmftop{o0,o1,o2,o3,o4,o5,o6}
				\fmfbottom{u0,u1,u2,u3,u4,u5,u6}
				\fmf{phantom}{u1,v1,o3}
				\fmf{plain}{v1,o3}
				\fmf{phantom}{o1,v1,u3}
				\fmf{plain}{v1,u3}
				\fmf{phantom}{u3,v2,o5}
				\fmf{plain}{u3,v2}
				\fmf{phantom}{o3,v2,u5}
				\fmf{plain}{o3,v2}
				\fmf{phantom}{u0,v1,dummy1,o4}
				\fmf{plain}{v1,dummy1}
				\fmf{phantom}{u4,dummy2,v1,o0}
				\fmf{plain}{dummy2,v1}
				\fmf{phantom}{u2,dummy2,v2,o6}
				\fmf{plain}{dummy2,v2}
				\fmf{phantom}{u6,v2,dummy1,o2}
				\fmf{plain}{v2,dummy1}
				\fmfv{decor.shape=circle,decor.filled=empty, decor.size=6.5thin}{v1,v2}
			\end{fmfgraph*}
		} + \dots	
	\end{eqnarray}
\end{fmffile}
Their symmetry factors are rather easy to determine: in the diagram of order
$n$, $n-1$ vertex flips generate a new labeled diagram, so that
one factor $\frac{1}{2}$ remains and in total the prefactor reads $\frac{1}{2}\frac{1}{n!}$.
The cumulants of the unperturbed theory being identical, we can resum these contributions, obtaining
\begin{equation}
	\sum_{n=1}^{\infty}\frac{1}{2}\frac{\left[w\left(\boldsymbol{x},\boldsymbol{y}\right)\right]^{n}}{n!}=\frac{1}{2}\left(\exp\left(w\left(\boldsymbol{x},\boldsymbol{y}\right)\right)-1\right)=:\frac{1}{2}f\left(\boldsymbol{x},\boldsymbol{y}\right),\label{eq:Def_Mayer_fct}
\end{equation}
where $f$ is known as Mayer function. What is now left to be shown is that we can perform this resummation also with diagrams with more than two nodes. We therefore conjecture that we obtain the full perturbation expansion of $\Gamma_{1}$ by the following rules:
\begin{enumerate}
	\item Draw all one-particle irreducible diagrams with each pair of nodes connected by at most one interaction line\label{Diagrammatic_rule_I_simple_liquids}
	\item Determine the symmetry factor of the diagram as usual
	\item Assign a factor $f\left(\boldsymbol{x},\boldsymbol{y}\right)$ to a line connecting nodes representing the densities $\rho\left(\boldsymbol{x}\right)$ and $\rho\left(\boldsymbol{y}\right)$ and integrate over all variables\label{Diagrammatic_rule_III_simple_liquids}.
\end{enumerate}	 
First, consider an interaction edge that is distinguished in the topology of a given diagram, like the one connecting the opposite nodes in
\begin{eqnarray}
	\begin{fmffile}{One-diagonal_diagram}	
		\parbox{25mm}{
			\begin{fmfgraph*}(75,25)
				\fmfpen{0.5thin}
				\fmftop{o1,o2,o3}
				\fmfbottom{u1,u2,u3}
				\fmf{phantom}{u1,dul,dml,v1,o2}
				\fmf{plain}{dml,v1,o2}
				\fmf{phantom}{u2,v2,dml,dol,o1}
				\fmf{plain}{u2,v2,dml}
				\fmf{phantom}{u2,v3,dmr,dor,o3}
				\fmf{plain}{u2,v3,dmr}
				\fmf{phantom}{u3,dur,dmr,v4,o2}
				\fmf{plain}{dmr,v4,o2}
				\fmf{plain, tension=0.}{v1,v3}
				\fmfv{decor.shape=circle,decor.filled=empty, decor.size=6.5thin}{v1,v2,v3,v4}
			\end{fmfgraph*}
		}
	\end{fmffile}\label{One-diagonal_diagram_eq}
\end{eqnarray}
Assume that in the remainder of the diagram, with the symmetry factor $S_{\mathrm{rest}}$, there are $k_{\mathrm{rest}}$ interactions. We replace the distinguished interaction by a $k$-fold interaction, which yields a new diagram with the prefactor
\[
	\fl \frac{1}{\left(k+k_{\mathrm{rest}}\right)!}\left(\begin{array}{c} k+k_{\mathrm{rest}}\\
	k
	\end{array}\right)S_{\mathrm{rest}}=\frac{1}{k!}\frac{S_{\mathrm{rest}}}{k_{\mathrm{rest}}!},
\]
coming about because we have $\left(\begin{array}{c} k+k_{\mathrm{rest}}\\ k \end{array}\right)$ possibilities to pick $k$ out of $k+k_{\mathrm{rest}}$ interactions to be part of the $k$-fold connection. Therefore, the only difference between the prefactor of the  enhanced diagram to that of the original one is indeed $1/k!$. 

Most interactions, however, are not topologically distinguished in a given diagram. By replacing one of them by a $k$-fold interaction, we distinguish it from the others and therefore, we potentially change the symmetry factor of the whole diagram. We have to make sure that this does not spoil the possibility of a resummation of interactions in Mayer functions. Assume that we have $N$ equivalent interactions (like the four outer interactions in (\ref{One-diagonal_diagram_eq})). We now choose an arbitrary partition
\[
	\sum_{i}N_{i}=N
\]
of this number and replace $N_{i}$ of the interaction lines by $k_{i}$-fold
ones (with $k_{i}=1,2,\dots$). Now, the prefactor of this (part of
the) diagram is given by
\begin{equation}
	\frac{1}{\left(\sum_{i}k_{i}N_{i}\right)!}\frac{\left(\sum_{i}k_{i}N_{i}\right)!}{\prod_{i}\left(k_{i}!\right)^{N_{i}}N_{i}!}=\frac{1}{\prod_{i}\left(k_{i}!\right)^{N_{i}}N_{i}!},	\label{eq:prefactor_equivalent_interactions}
\end{equation}
because we have $\left(\begin{array}{c} \sum_{i}k_{i}N_{i}\\ k_{1} \end{array}\right)$ possibilities to choose interactions for the first group of multiple connections, $\left(\begin{array}{c}\sum_{i}k_{i}N_{i}-k_{1}\\ k_{1}
\end{array}\right)$ for the second (if $N_{1}>1$) and so on. Because we have $N_{i}$
equivalent spots for interactions with $k_{i}$-fold interactions, we have to divide this expression by $N_{i}!$ in addition. 
Let's show that this yields the same expression as using the rules with the resummed interactions. To this end, using the extended diagrammatic rule conjectured before, we insert the resummation Eq.~\prettyref{eq:Def_Mayer_fct} into the expression for the diagram with single interactions, which bears $\frac{1}{N!}$ as prefactor because the diagram without resummations contains $N$ interactions.
Multiplying out the product of the series expansions of the exponential functions, we
observe that there are $\frac{N!}{\prod_{i}N_{i}!}$ possibilities to choose $N_{i}$ monomials with degree $k_{i}$ out of $N$ factors for all $i$. The factor $\frac{1}{\prod_{i}\left(k_{i}!\right)^{N_{i}}}$
is contained in the definition of the exponential function.\\
We therefore recover the same prefactor Eq.~\prettyref{eq:prefactor_equivalent_interactions} by applying the foreshadowed rules as by resumming the original diagrams. This completes the proof that we can replace the standard diagrammatic rules by (\ref{Diagrammatic_rule_I_simple_liquids})-(\ref{Diagrammatic_rule_III_simple_liquids}). Because multiple connecting lines do not occur here, interactions are normally represented by straight lines in the simple-liquids literature; we, however, do not stick to this convention to avoid confusion with propagator lines of Gaussian theories. 

\subsubsection{Second Mayer theorem}\label{sub:Second_Mayer_thm}

Now, we change our variable to describe the system's state: We do not use the source term (or single-particle potential) $\nu$, but its conjugate variable, the (local) density $\rho$, by using the
(first) Legendre transform to define
\[
	\Gamma_{1}\left[\rho,w\right]=\sup_{\nu}\left(\rho\nu-W\left[\nu,w\right]\right)
\]
To derive its diagrammatic representation, we proceed as for the cumulant-generating functional, that is we write
\begin{eqnarray}
	\fl e^{-\Gamma_{1}^{V}\left[\rho,w\right]} & =e^{-W_{0}\left[\nu\right]}\left(1-\frac{\beta}{L}A_{1}\left[\frac{\delta}{\delta\nu}\right]\right)^{l}e^{W_{0}\left[\nu\right]},\\
	\fl \mathrm{where}\ A_{1}\left[\frac{\delta}{\delta\nu}\right] & =-\frac{1}{2}\int d\boldsymbol{x}\int d\boldsymbol{y}\,w\left(\boldsymbol{x},\boldsymbol{y}\right)\left[\frac{\delta}{\delta\nu\left(\boldsymbol{x}\right)}\frac{\delta}{\delta\nu\left(\boldsymbol{y}\right)}-\delta\left(\boldsymbol{x}-\boldsymbol{y}\right)\frac{\delta}{\delta\nu\left(\boldsymbol{x}\right)}\right]\label{eq:A_operator_contr_interaction}\\
 	\fl & +\int d\boldsymbol{x}\frac{\delta\Gamma_{1}^{V}}{\delta\rho\left(\boldsymbol{x}\right)}\left(\frac{\delta}{\delta\nu\left(\boldsymbol{x}\right)}-\rho\left(\boldsymbol{x}\right)\right).\label{eq:A_operator_contr_Gamma_prime}
\end{eqnarray}
This is a special case of Eq.~(12) in \cite{Kuehn18_375004}, with $j$ and $x$ relabeled as $\nu$ and $\rho$, respectively. Note that there, summations over indices and integration over continuous variables are implied (here we made this explicit by writing out the integrals over the positions of the particles). 
Constructing the diagrams, we can now argue exactly as in \cite{Kuehn18_375004}, up to the fact that, in the present case, all cumulants of the unperturbed theory, \prettyref{eq:cumulants_ideal_gas}, are equal. Therefore, we do not need to use the relation $\Gamma_{V}\left[\rho\right]=-g\left[\Gamma_{0}^{\left(1\right)}\left[\nu\right]\right]$ to determine the derivative of $\Gamma_{V}$, but we can evaluate it directly: $\frac{\delta}{\delta \rho}\Gamma_{1}^{\ii}$ is represented by all diagrams of $\Gamma_{1}^{\ii}$ with one node removed in every possible way. Computing the symmetry factor of the entire diagram, we argue as in the Gaussian case that diagrams of a given shape always bear the same symmetry factor, no matter if they were generated by \prettyref{eq:A_operator_contr_interaction} or \prettyref{eq:A_operator_contr_Gamma_prime} - however, they occur with opposite signs.\\
Therefore, in a way similar to perturbations around Gaussian theories, all diagrams that decompose into two pieces by removing a cumulant get canceled: only one-particle irreducible diagrams remain, which is known as second Mayer theorem. This is a peculiarity that the unperturbed Poisson and Gaussian distributions share in common (for other non-Gaussian theories, a new class of diagrams would have to be introduced for the first Legendre transform~\cite{Kuehn18_375004}).

\subsection{Some identities needed for the diagrammatic expansion of the simple liquid}
\subsubsection{For isolated ring diagrams}\label{Number_theory_isolated_ring}
For the construction of ring diagrams, we need the following identity:
\begin{eqnarray}
	\label{eq:identity_ring_diagrams}
	\fl \frac{\left(-1\right)^{n+1}}{n} &= \sum_{\boldsymbol{k}\in P_{r}\left(n\right)}\left(\sum_{i=1}^{n-2}k_{i}-1\right)!\prod_{i=1}	^{n-2}\frac{\left(-1\right)^{\left(i+1\right)k_{i}}}{k_{i}!} \\
	\fl &= \sum_{\boldsymbol{k}\in P\left(n\right)}\left(\sum_{i=1}^{n}k_{i}-1\right)!\prod_{i=1}^{n}\frac{\left(-1\right)^{\left(i+1\right)k_{i}}}{k_{i}!}\, \forall n\geq 3,\label{eq:number_theory_ring_diagrams_second_line}
\end{eqnarray}
where $P_{r}\left(n\right)$ is defined in Eq.~\prettyref{eq:Def_reduced_partition_fct}.
In \prettyref{eq:number_theory_ring_diagrams_second_line}, we have extended the range of the $\boldsymbol{k}$-sum from $P_{r}\left(n\right)$ to $P\left(n\right)$. This is possible because it only adds the entries with $k_{1}=k_{n-1}=1,\ k_{i}=0\,\forall i\neq1,n-1$ and $k_{n}=1,\ k_{i}\neq\,\forall i\neq n$, yielding contributions $1$ and $-1$ in \prettyref{eq:identity_ring_diagrams}, so they cancel. We can go a step further and let the dimension of $\boldsymbol{k}$ tend to infinity because $k_{i}$ will be 0 automatically for $i>n$:
\begin{eqnarray}
	\fl &\sum_{\boldsymbol{k}\in P\left(n\right)}\left(\sum_{i=1}^{n}k_{i}-1\right)!\prod_{i=1}^{n}\frac{\left(-1\right)^{\left(i+1\right)k_{i}}}{k_{i}!} \\
	\label{eq:identity_ring_diagrams_infty}
	 \fl =  &\lim_{m\rightarrow\infty}\sum_{k_{1},\dots,k_{n}=0,\,\sum_{i=1}^{m}k_{i}\neq0}^{\infty} \delta_{n,\sum_{i=1}^{n} i k_{i}} \left(\sum_{i=1}^{m}k_{i}-1\right)!\prod_{i=1}^{m}\frac{\left(-1\right)^{\left(i+1\right)k_{i}}}{k_{i}!}
\end{eqnarray}
We multiply \prettyref{eq:identity_ring_diagrams_infty} with $z^{n}$ (with $z\in\mathbb{R}, \left|z\right|<1$) and sum $n$ from $1$ to $\infty$ (we only need it from $n=3$ on, but it gets easier like this):
\begin{eqnarray}
	\fl Z_{r}\left(z\right) & :=  \sum_{n=1}^{\infty}z^{n} \lim_{m\rightarrow\infty} \sum_{k_{1},\dots,k_{n}=0,\,\sum_{i=1}^{m}k_{i}\neq0}^{\infty} \delta_{n,\sum_{i=1}^{n} i k_{i}} \left(\sum_{i=1}^{m}k_{i}-1\right)!\prod_{i=1}^{m}\frac{\left(-1\right)^{\left(i+1\right)k_{i}}}{\prod_{i=1}^{n}k_{i}!} \\
	\fl &= \lim_{m\rightarrow\infty}\sum_{k_{1},\dots,k_{n}=0,\,\sum_{i=1}^{m}k_{i}\neq0}^{\infty}\left(\sum_{i=1}^{m}k_{i}-1\right)!\prod_{i=1}^{m}\left(\frac{\left(-1\right)^{\left(i+1\right)k_{i}}}{k_{i}!}z^{k_{i}\cdot i}\right).
\end{eqnarray}
We represent the factorial of the sum as $\left(\sum_{i=1}^{m}k_{i}-1\right)!=\int_{0}^{\infty}e^{-u}u^{\sum_{i=1}^{m}k_{i}-1}du$. This integral expression can also be written down for $\sum_{i=1}^{m}k_{i}=0$ so that we can lift the remaining restriction on the $\boldsymbol{k}$-sum and instead substract the contribution for $\sum_{i=1}^{m}k_{i}=0$:
\begin{eqnarray}
	\fl Z_{r}\left(z\right) &= \lim_{m\rightarrow\infty} \sum_{k_{1},\dots,k_{n}=0}^{\infty}\int_{0}^{\infty}e^{-u}u^{\sum_{i=1}^{m}k_{i}-1}du\prod_{i=1}^{m}\left(\frac{\left(-1\right)^{\left(i+1\right)k_{i}}}{\prod_{i=1}^{n}k_{i}!}z^{k_{i}\cdot i}\right)-\int_{0}^{\infty}\frac{e^{-u}}{u}du \label{eq:P_ring_diagrams_integral_rep_factorial}\\
	\fl = &\lim_{m\rightarrow\infty} \sum_{k_{1},\dots,k_{n}=0}^{\infty}\int_{0}^{\infty}\frac{e^{-u}}{u}\prod_{i=1}^{m}\left(\frac{\left(-1\right)^{\left(i+1\right)k_{i}}}{k_{i}!}z^{k_{i}\cdot i}u^{k_{i}}\right)du-\int_{0}^{\infty}\frac{e^{-u}}{u}du \\
	\fl &= \int_{0}^{\infty}\frac{e^{-u}}{u}\left[-1+\lim_{m\rightarrow\infty}\prod_{i=1}^{m}\sum_{k_{i}=0}^{\infty}\frac{\left[\left(-1\right)^{\left(i+1\right)}z^{i}u\right]^{k_{i}}}{k_{i}!}\right] \\
	\fl &= \int_{0}^{\infty}\frac{e^{-u}}{u}\left[-1+\lim_{m\rightarrow\infty}\prod_{i=1}^{m}\exp\left(-\left(-z\right)^{i}u\right)\right] \\
	\fl &= \int_{0}^{\infty}\frac{e^{-u}}{u}\left[-1+\exp\left(-u\lim_{m\rightarrow\infty}\sum_{i=1}^{m}\left(-z\right)^{i}\right)\right]
\end{eqnarray}		
We use that, because $\left|z\right|<1$, $\lim_{m\rightarrow\infty}\sum_{i=1}^{m}\left(-z\right)^{i}=\lim_{m\rightarrow\infty}\frac{z^{m+1}-z}{z+1}=-\frac{z}{1+z}$ to obtain
\begin{eqnarray}
	\fl Z_{r}\left(z\right) &=\int_{0}^{\infty}\frac{-\exp\left(-u\right)+\exp\left(u\left(-1+\frac{z}{z+1}\right)\right)}{u}du\\
	\fl &=\int_{0}^{\infty}\frac{\exp\left(u\left(-\frac{1}{z+1}\right)\right)-\exp\left(-u\right)}{u}du \label{eq:Partition_fct_ring_nearly_solved}\\
	\fl &=\int_{0}^{\infty}\frac{\exp\left(u\right)-\exp\left(-\left(z+1\right)u\right)}{u}du,
\end{eqnarray}
therefore
\begin{eqnarray}
	\fl \frac{d}{dz}Z_{r}\left(z\right) &= -\int_{0}^{\infty}\exp\left(-u\right)\frac{\frac{\partial}{\partial z}\exp\left(-zu\right)}{u}du\\
	\fl &=\int_{0}^{\infty}\exp\left(-\left(z+1\right)u\right)du\\
	\fl &=-\frac{1}{z+1}\left.\exp\left(-\left(z+1\right)u\right)\right|_{0}^{\infty}=\frac{1}{z+1}.
\end{eqnarray}
Together with $Z_{r}\left(0\right)=0=\ln\left(1+0\right)$ and $\frac{d}{dz}\ln\left(1+z\right)=\frac{1}{1+z}$, this proves that
\begin{eqnarray}
	\fl Z_{r}\left(z\right)=\ln\left(1+z\right)=\sum_{n=1}^{\infty}\left(-1\right)^{n+1}\frac{z^{n}}{n}.
\end{eqnarray}	
So, by comparing the coefficients of $n$, we obtain the desired result, Eq.~\prettyref{eq:identity_ring_diagrams}.

\subsubsection{Identity for caterpillars integrated in two-particle reducible diagrams}\label{Number_theory_caterpillars}
We now would like to derive that
\begin{eqnarray}
	\fl 0 = \sum_{\boldsymbol{k}\in P\left(n\right)}\left(\sum_{i=1}^{n}k_{i}\right)!\prod_{i=1}^{n}\frac{\left(-1\right)^{\left(i+1\right)k_{i}}}{k_{i}!} \ \forall n\neq 1.
\end{eqnarray}
The only difference to the claim in the preceeding section is the factor $\left(\sum_{i=1}^{n}k_{i}\right)!$ replacing $\left(\sum_{i=1}^{n}k_{i}-1\right)!$. Therefore, we just have to remove the $\frac{1}{u}$ in Eq.~\prettyref{eq:P_ring_diagrams_integral_rep_factorial} and the following ones, in particular Eq.~\prettyref{eq:Partition_fct_ring_nearly_solved} to obtain for the corresponding partition function 
\begin{eqnarray}
	\fl Z_{ca}\left(z\right) &= \int_{0}^{\infty}\exp\left(u\left(-\frac{1}{z+1}\right)\right)-\exp\left(-u\right)du \\
	\fl &=-\left(z+1\right)\left.\exp\left(u\left(-\frac{1}{z+1}\right)\right)\right|_{0}^{\infty}+\left.\exp\left(-u\right)\right|_{0}^{\infty}\\
	\fl & =z+1-1=z,
\end{eqnarray}
so all coefficients of $P_{rc}\left(z\right)$ but the linear one vanish.

\subsubsection{Identity for multiple connections in two-particle reducible diagrams}\label{Number_theory_multiple_connections}
We would like to demonstrate that
\begin{eqnarray}
	\label{eq:Statement_cherry_combined}
	\fl 0 &= \sum_{\boldsymbol{k}\in P\left(n\right)}\prod_{i=1}^{n}\frac{\left(-1\right)^{\left(i+1\right)k_{i}}}{k_{i}!i^{k_{i}}}\\
	\fl  &= \lim_{m\rightarrow\infty}\sum_{k_{1}=0,\dots,k_{m}=0,\,\sum_{i=1}^{m}k_{i}\neq0}^{\infty}\delta_{\sum_{i=1}^{m}k_{i}\cdot i,n}\prod_{i=1}^{m}\frac{\left(-1\right)^{\left(i+1\right)k_{i}}}{k_{i}!i^{k_{i}}} \ \forall n\neq 1.
\end{eqnarray}
With the same trick as in appendix \ref{Number_theory_isolated_ring}, i.e. multiplying by $z^{n}$, summing over $n$ from $1$ to $\infty$, exchanging the order of summation, performing the sum over $n$ and finally adding and deducing the contribution from $\sum_{i=1}^{m}k_{i}=0$ (which yields $1$), we obtain for $\left|z\right|<1$
\begin{eqnarray}
	\fl Z_{mc}\left(z\right) &:= \sum_{n=1}^{\infty}z^{n}\lim_{m\rightarrow\infty}\sum_{k_{1}=0,\dots,k_{m}=0,\,\sum_{i=1}^{m}k_{i}\neq0}^{\infty}\delta_{\sum_{i=1}^{m}k_{i}\cdot i,n}\prod_{i=1}^{m}\frac{\left(-1\right)^{\left(i+1\right)k_{i}}}{k_{i}!i^{k_{i}}} \\
	\fl &= \lim_{m\rightarrow\infty}\sum_{k_{1}=0,\dots,k_{m}=0}^{\infty}\prod_{i=1}^{m}\frac{\left(-1\right)^{\left(i+1\right)k_{i}}}{k_{i}!i^{k_{i}}}z^{\sum_{i=1}^{m}k_{i}\cdot i}-1 \\
	\fl &= \lim_{m\rightarrow\infty}\prod_{i=1}^{m}\sum_{k_{i}=0}^{\infty}\frac{\left(-1\right)^{\left(i+1\right)k_{i}}\left(z^{i}\right)^{k_{i}}}{k_{i}!i^{k_{i}}}-1\\
	\fl &= \lim_{m\rightarrow\infty}\prod_{i=1}^{m}\left(\exp\left(\left(-1\right)^{\left(i+1\right)}\frac{z^{i}}{i}\right)\right)-1\\
	\fl &= \lim_{m\rightarrow\infty}\exp\left(-\sum_{i=1}^{m}\left(-1\right)^{i}\frac{z^{i}}{i}\right)-1.
\end{eqnarray}	
We use $\sum_{i=1}^{m}\frac{\left(-z\right)^{i}}{i}=\sum_{i=1}^{m}\int_{0}^{-z}t^{i-1}dt=\int_{0}^{-z}\sum_{i=1}^{m}t^{i-1}dt=\int_{0}^{-z}\frac{1-t^{m}}{1-t}dt$ to obtain
\begin{eqnarray}
	\fl Z_{mc}\left(z\right) &= \exp\left(-\lim_{m\rightarrow\infty}\int_{0}^{-z}\frac{1-t^{m}}{1-t}dt\right)-1 = \exp\left(-\int_{0}^{-z}\frac{1}{1-t}dt\right)-1  \\
	\fl &= \exp\left(\left.\ln\left(1-t\right)\right|_{0}^{-z}\right)-1 = \exp\left(\ln\left(1+z\right)\right)-1 \\
	\fl &=z,
\end{eqnarray}	
where the integrand in the first line is uniformly convergent so that we can exchange the order of performing the limit and the integration. Comparing the coefficients of the partition function, we obtain \prettyref{eq:Statement_cherry_combined}.

\subsubsection{Identity for isolated watermelon diagrams}\label{Number_theory_isolated_cherry}
Finally, we would like to prove that
\begin{eqnarray}
	\sum_{\boldsymbol{k}\in P_{r}\left(n\right)}\prod_{i=1}^{n-2}\frac{\left(-1\right)^{\left(i+1\right)k_{i}}}{k_{i}!i^{k_{i}}}=\left(-1\right)^{n+1}\frac{1}{n\left(n-1\right)}\ \forall n\geq 3. \label{eq:number_theory_cherry_diagrams}
\end{eqnarray}
We observe that
\begin{eqnarray}
	\fl &\sum_{\boldsymbol{k}\in P_{r}\left(n\right)}\prod_{i=1}^{n-2}\frac{\left(-1\right)^{\left(i+1\right)k_{i}}}{k_{i}!i^{k_{i}}} \\	
	=&\sum_{\boldsymbol{k}\in P\left(n\right),\,\sum_{i=1}^{n}k_{i}\neq0}\prod_{i=1}^{n}\frac{\left(-1\right)^{\left(i+1\right)k_{i}}}{k_{i}!i^{k_{i}}}-\left(\frac{\left(-1\right)^{n}}{n-1}+\frac{\left(-1\right)^{n+1}}{n}\right)\,\forall n\geq 3,
\end{eqnarray}
where the compensating terms come from the cases $k_{n-1}=k_{1}=1,\ k_{i}=0\,\forall i\neq1,n-1$ and $k_{n}=1,\ k_{i}=0\,\forall i\neq n$. So we define the partition function by 
\begin{eqnarray}
	\fl Z_{ch}\left(z\right) &:=  \sum_{n=3}^{\infty}z^{n}\left[\sum_{\boldsymbol{k}\in P\left(n\right),\,\sum_{i=1}^{n}k_{i}\neq0}\prod_{i=1}^{n}\frac{\left(-1\right)^{\left(i+1\right)k_{i}}}{k_{i}!i^{k_{i}}}-\left(\frac{\left(-1\right)^{n}}{n-1}+\frac{\left(-1\right)^{n+1}}{n}\right)\right]\\
	\fl &=  -z+Z_{mc}\left(z\right)+\sum_{n=2}^{\infty}z^{n}\left(-1\right)^{n}\left(-\frac{1}{n-1}+\frac{1}{n}\right)\\
	\fl &=  -\sum_{n=3}^{\infty}z^{n}\left(-1\right)^{n}\frac{1}{n\left(n-1\right)},
\end{eqnarray}
where we have used that the second-order term in the first part of the sums cancels and have indentified the rest as $Z_{mc}$. This proves \prettyref{eq:number_theory_cherry_diagrams}.\\


\providecommand{\newblock}{}

\end{document}